\documentclass[11pt,fullpage]{article}
\usepackage{geometry}		
\usepackage[parfill]{parskip}	
\usepackage{graphicx}
\usepackage{amssymb}
\usepackage{epsfig}
\usepackage{epstopdf}
\usepackage[scaled]{helvet}
\usepackage{amsmath,array}
\usepackage{amssymb,amscd,amsfonts}
\usepackage{nameref}
\usepackage{thumbpdf}
\usepackage{subfigure}
\usepackage{multirow}

\newcommand{\be}{\begin{equation}}
\newcommand{\ee}{\end{equation}}
\newcommand{\bea}{\begin{eqnarray}}
\newcommand{\eea}{\end{eqnarray}}

\renewcommand{\vec}[1]{{ {\bf #1  }}}

\renewcommand{\r}[1]{\mathrm{#1}}
\newcommand{\bi}{\begin{itemize}}
\newcommand{\ei}{\end{itemize}}
\newcommand{\ben}{\begin{enumerate}}
\newcommand{\een}{\end{enumerate}}

\newcommand{\md}{\mathrm{d}}	

\newcommand{\bcen}{\begin{center}}
\newcommand{\ecen}{\end{center}}

\newcommand{\bmc}{\begin{multicols}{2}}
\newcommand{\emc}{\end{multicols}}

\newcommand{\bsym}{\boldsymbol}

\newcommand{\rev}[1]{#1}

\DeclareGraphicsExtensions{.png,.gif,.jpg,.pdf,.bmp}

\title{A new class of highly efficient exact stochastic simulation algorithms for chemical reaction networks}
\author{Rajesh Ramaswamy, N\'elido Gonz\'alez-Segredo 
and Ivo F.~Sbalzarini\\
rajeshr, nelidog, ivos@ethz.ch\\
\it{Institute of Theoretical Computer Science and Swiss Institute of Bioinformatics}\\
\it{ETH Zurich, CH--8092 Z\"urich, Switzerland}}
\date{\today}                                           
\begin{document}
\maketitle
\begin{abstract}
We introduce an alternative formulation of the exact stochastic simulation algorithm (SSA) for sampling trajectories of the chemical master equation for a well-stirred system of coupled chemical reactions. Our formulation is based on factored-out, partial reaction propensities. This novel exact SSA, called the partial propensity direct method (PDM), is highly efficient and has a computational cost that scales at most linearly with the number of chemical species, irrespective of the degree of coupling of the reaction network. In addition, we propose a sorting variant, SPDM, which is especially efficient for multiscale reaction networks.\\
\\
\\
\textit{\textbf{This article has been accepted by The Journal of Chemical Physics. After it is published, it will be found at http://jcp.aip.org/}.}
\end{abstract}

\section{Introduction}

In chemical kinetics, the temporal evolution of a well-stirred system of chemically reacting molecules is classically described using reaction rate equations. Reaction-rate equations are a mean-field description formulated as coupled ordinary differential equations. The number of molecules is continuous in time, and reaction rates are quantified using macroscopic rate constants. Such reaction rate equations, however, do not always provide an accurate description. This is the case especially, but not only, when the number of molecules of the various chemical species (henceforth called the population) is much smaller than Avogadro's number~\cite{Qian:2002,Shibata:2004}. At low population, the number of molecules is not large enough for fluctuations to be negligible. In addition, fluctuations may play an important role in the kinetics~\cite{Qian:2002,Li:2008a}. Even at high population, correlated fluctuations can cause the mean to behave in a way that is not captured by a mean-field description~\cite{Shibata:2004,Gardiner:1976,Engblom:2006}. These effects can be accounted for by stochastic kinetic models, which can incorporate thermal fluctuations in a number of ways. An approach that has become canonical is the chemical master equation (CME)~\cite{Gillespie:1976,Gillespie:1977,Gillespie:1992}, a Markov-chain model with many applications in physics, chemistry, and biology. The CME models the kinetics of any chemical reaction system that is well stirred and thermally equilibrated \cite{Gillespie:1992}. Its high dimensionality, however, renders analytical approaches intractable. \\

Numerical methods to sample trajectories from the CME mostly rely on kinetic Monte Carlo approaches~\cite{Bortz:1975}. The canonical kinetic Monte Carlo approach for sampling a trajectory of the CME is Gillespie's stochastic simulation algorithm (SSA)~\cite{Gillespie:1976,Gillespie:1977,Gillespie:1992}. SSA is governed by the joint probability density 
\be
p(\tau,\mu | \vec{n}(t)) = (a \text{e}^{-a\tau})( a_{\mu}/a)
\label{eq:jointpdf}
\ee
for the random variables $\tau$ (the time to the next reaction) and $\mu$ (the index of the next reaction). The vector \mbox{$\vec{n}(t)=(n_{1},\dots,n_{N})$} is the population at time $t$. Each entry \mbox{$n_i$} is the number of molecules of the respective species S\mbox{$_i$}, and $N$ is the total number of species. The propensity of each reaction $\mu$ is defined as \mbox{$a_\mu = c_\mu h_\mu$}, where \mbox{$c_{\mu}$} is the specific probability rate, and \mbox{$h_{\mu}= h_{\mu}(\vec{n})$} is the reaction degeneracy, which is the number of possible combinations of reactants in reaction $\mu$ given the population $\vec{n}$. 
The reaction propensity is such that \mbox{$a_{\mu}\md t$} is the probability that a randomly selected combination of reactant molecules of reaction $\mu$ at time $t$ will react in the next infinitesimal time interval \mbox{$\md t$}. The total propensity is \mbox{$a=\sum_{\mu=1}^{M}a_{\mu}$}, where $M$ is the total number of reactions. \\

Existing SSA formulations can be classified into {\em exact} and {\em approximate} methods. Exact methods sample from the probability density in Eq.~\ref{eq:jointpdf}. These formulations include the direct method (DM)~\cite{Gillespie:1976,Gillespie:1992}, the first reaction method (FRM)~\cite{Gillespie:1976}, Gibson-Bruck's next-reaction method (NRM)~\cite{Gibson:2000}, a Gibson-Bruck variant of the DM~\cite{Gibson:2000}, the optimized direct method (ODM)~\cite{Cao:2004}, the sorting direct method (SDM)~\cite{McCollum:2006}, the logarithmic direct method (LDM) (unpublished, \cite{Li:2006b}), and the composition-rejection formulation (SSA-CR) \cite{Slepoy:2008}. 
Approximate SSA formulations provide better computational efficiency for large numbers of molecules by sampling from an approximation to the probability density in Eq.~\ref{eq:jointpdf}. These methods include $\tau$-leaping~\cite{Gillespie:2001,Cao:2005,Cao:2006,Peng:2007a}, $k_\alpha$-leaping~\cite{Gillespie:2001}, $R$-leaping~\cite{Auger:2006a}, $L$-leap~\cite{Peng:2007}, $K$-leap~\cite{Cai:2007}, the slow-scale method~\cite{Cao:2005a}, and implicit $\tau$-leaping~\cite{Rathinam:2003}.\\ 

In this paper we focus on exact methods. They offer the advantage of being parameter-free, whereas all approximate methods contain parameters that need to be adjusted by the user. The computational cost of exact SSA formulations is dominated by the steps needed to sample the next reaction and update the propensities after a reaction has fired \cite{Gibson:2000,Cao:2004,Slepoy:2008}. In Gillespie's original DM and FRM, this leads to a computational cost that scales linearly with the number of reactions in the network. Various improved SSA formulations have been proposed in order to reduce this computational cost. 
The most notable improvements include the use of dependency graphs to reduce the number of propensities that need to be updated \cite{Gibson:2000}, and various sampling schemes of higher efficiency \cite{Gibson:2000,Cao:2004,McCollum:2006,Slepoy:2008}. All of these sampling schemes can be interpreted as instances of the random-variate generation problem \cite{Slepoy:2008} as described in Devroye's compendium \cite{Devroye:1986} and can reduce the computational cost (CPU time) of sampling the next reaction. These improvements have reduced the computational cost of SSA to logarithmic or even constant scaling for weakly coupled networks. 
For strongly coupled networks, however, the computational cost of all improved SSA formulations still scales linearly with the number of reactions. We define {\em weakly coupled} networks as those where the maximum number of reactions that are influenced by any other reaction, i.e.~the maximum degree of coupling of the network, is independent of system size. This is in contrast to {\em strongly coupled} networks, where the number of influenced reactions grows proportionally with system size and can be as large as the total number of reactions. In such networks, the total number of reactions grows faster than the number of species when the latter is increased. Strongly coupled networks frequently occur, e.g., in nucleation-and-growth models, scale-free biochemical networks, and colloidal aggregation systems. In these cases, the scaling of the computational cost of most of the improved SSA formulations with system size is equivalent to that of DM (see Sec.~\ref{sec:oldssa}). \\

We present a novel SSA formulation with a computational cost that scales at most linearly with the number of {\em species}, making it especially efficient for strongly coupled networks. This is made possible by restricting the class of systems to networks containing only elementary chemical reactions, where every reaction has at most two reactants \cite{Gillespie:1992}. This allows factoring out one of the species from every reaction propensity, leading to {\em partial propensities} that depend on the population of at most one species. Any non-elementary reaction can always be broken down into elementary reactions, at the expense of an increase in system size \cite{Gillespie:1992,Wilhelm:2000,Schneider:2000}. The use of partial propensities leads to SSA formulations with a computational cost that scales as some function of the number of species rather than the number of reactions. \\

In Sec.~\ref{sec:methods_description}, we formally introduce the concept of partial propensities and present two partial propensity variants of the exact SSA: the partial-propensity direct method (PDM) and the sorting partial-propensity direct method (SPDM). They use partial propensities and efficient data structures for sampling the next reaction and for updating the partial propensities after a reaction. We benchmark them in Sec.~\ref{sec:benchmarks} and show that their computational cost scales at most linearly with the number of species in the network, irrespective of the degree of coupling. The benchmarks include two strongly coupled networks, for which the degree of coupling grows with system size, a weakly coupled reaction network with a constant maximum degree of coupling, and a small, fixed-size biological multiscale (stiff) network. In order to test the competitiveness of our algorithm in cases where several other SSA formulations might be more efficient, we choose the most weakly coupled network possible, the linear chain model, where the number of reactions scales linearly (with a proportionality constant of 1) with the number of species \cite{Albert:2005,Cao:2004}. The multiscale biological network is included in order to benchmark the new algorithms on small systems and when the reaction propensities span several orders of magnitude. In Sec.~\ref{sec:conclusion} we summarize the main results, discuss the limitations of the presented method, and give an outlook on possible future developments and applications.

\section{Computational cost of previous exact SSA formulations}\label{sec:oldssa}

We review the scaling of the computational cost of previous exact SSA formulations. In order to express scaling with system size $x$, we use the Bachmann-Landau notation, writing $C(x)\in O(f(x))$ ($C(x)$ is $O(f(x))$) whenever $C(x)>0$ is bounded from above by $f(x)$ as $C(x)\le \alpha f(x)$, for all $x$ and some constant pre-factor \mbox{$\alpha>0$} that is independent of $x$. \\

Since DM and FRM form the basis for most exact SSA's, we first focus on these two. DM's computational cost is \mbox{$O(M)$}~\cite{Gillespie:1976,Gillespie:1992,Gibson:2000,Cao:2004,Slepoy:2008}, where $M$ is the total number of reactions (see also Appendix \ref{app:ssa}). In FRM, the sampling strategy for $\mu$ and $\tau$ is different (see Appendix \ref{app:ssa}). This, however, does not change the scaling of the computational cost of FRM, which remains \mbox{$O(M)$}. Since the FRM sampling strategy involves discarding \mbox{$M-1$} reaction times, its computational cost generally has a larger pre-factor than that of DM~\cite{Gillespie:1976,Gibson:2000,Cao:2004}. \\

NRM is an improvement over FRM in which the \mbox{$M-1$} unused reaction times are suitably reused, and data structures such as indexed priority queues and dependency graphs are introduced. The indexed priority queue, which is equivalent to a heap tree, is used to sort the \mbox{$\tau_i$}'s more efficiently; the dependency graph is a data structure that contains the indices of the reactions whose propensities are to be recomputed after a certain reaction $\mu$ has fired. This avoids having to recompute all \mbox{$a_\mu$}'s after every reaction. Each reaction is represented as a node in the dependency graph, and nodes $i$ and $j$ are connected by a directed edge if and only if the execution of reaction $i$ affects the propensity (through the population of reactants) of reaction $j$. These data structures, together with the reuse of reaction times, reduce the computational cost of NRM to \mbox{$O(k\log _2 M)$}, where $k$ is the out-degree of the dependency graph, that is, the degree of coupling of the reaction network. In strongly coupled networks, $k$ is a function of $M$ and is $O(M)$. The computational cost of NRM is thus \mbox{$O(M)$} for strongly coupled networks. Even for some weakly coupled networks, the computational cost of NRM has been empirically shown to be \mbox{$O(M)$}~\cite{Cao:2004}. This is due to the additional overhead, memory-access operations, and cache misses introduced by the complex data structures (indexed priority queue, dependency graph) of NRM. The scaling of the computational cost of the Gibson-Bruck variant of DM is equal to that of NRM, albeit with a larger pre-factor~\cite{Gibson:2000}. For weakly coupled networks where $k(M)$ is $O(1)$, independent of system size, the computational cost is further reduced to $O(1)$ in the SSA-CR formulation \cite{Slepoy:2008} under the assumption that the ratio of maximum to minimum propensity is bounded. For strongly coupled networks, where $k(M)$ is $O(M)$, the computational cost of SSA-CR is $O(M)$ \cite{Slepoy:2008,Schulze:2008}. \\

ODM is an improvement over DM where the reactions are sorted in descending order of firing frequency. This makes it more probable to find the next reaction close to the beginning of the list and, hence, reduces the search depth for finding the index of the next reaction using linear search. ODM estimates the firing frequencies of all reactions during a short pre-simulation run of about 5--10\% of the length of the entire simulation \cite{Cao:2004,McCollum:2006}. In order to reduce the cost of updating the propensities after a reaction has fired, ODM also uses a dependency graph. Irrespective of the degree of coupling, the computational cost of ODM is \mbox{$O(M)$}, which was confirmed in benchmarks \cite{Cao:2004}. SDM is a variant of ODM that does not use pre-simulation runs, but dynamically shifts up a reaction in the reaction list whenever it fires (``bubbling up'' the more frequent reactions). This further reduces the pre-factor of the computational cost of SDM compared to that of ODM, but the scaling remains \mbox{$O(M)$} \cite{McCollum:2006}. \\

LDM uses a binary search tree (recursive bisection) on an ordered linear list of cumulative sums of propensities to find the next reaction. This is reported to reduce the average search depth of this step to \mbox{$O(\log _2 M)$}~\cite{Li:2006b}. Irrespective of the degree of coupling, however, the update step is $O(M)$ since on average \mbox{$(M+1)/2$} sums of propensities need to be recomputed, rendering the computational cost of LDM \mbox{$O(M)$}. \\

In summary, the computational cost of previously reported exact SSA formulations is \mbox{$O(M)$} for strongly coupled networks. For weakly coupled networks, however, some are significantly more efficient and can be $O(\log _2 M)$ or even $O(1)$. 

\section{Partial-propensity methods}\label{sec:methods_description}

We introduce the concept of partial propensities for elementary reactions and use it to formulate two partial-propensity direct methods, PDM and SPDM, whose computational cost scales at most linearly with the number of species, even for strongly coupled networks. SPDM uses concepts from SDM \cite{McCollum:2006} to dynamically rearrange reactions, which reduces the average search depth for sampling the next reaction in a multiscale network. \\

\begin{figure}[htbp]
\begin{center}
\subfigure[(a)]{{\label{Data_structures}}\epsfig{file=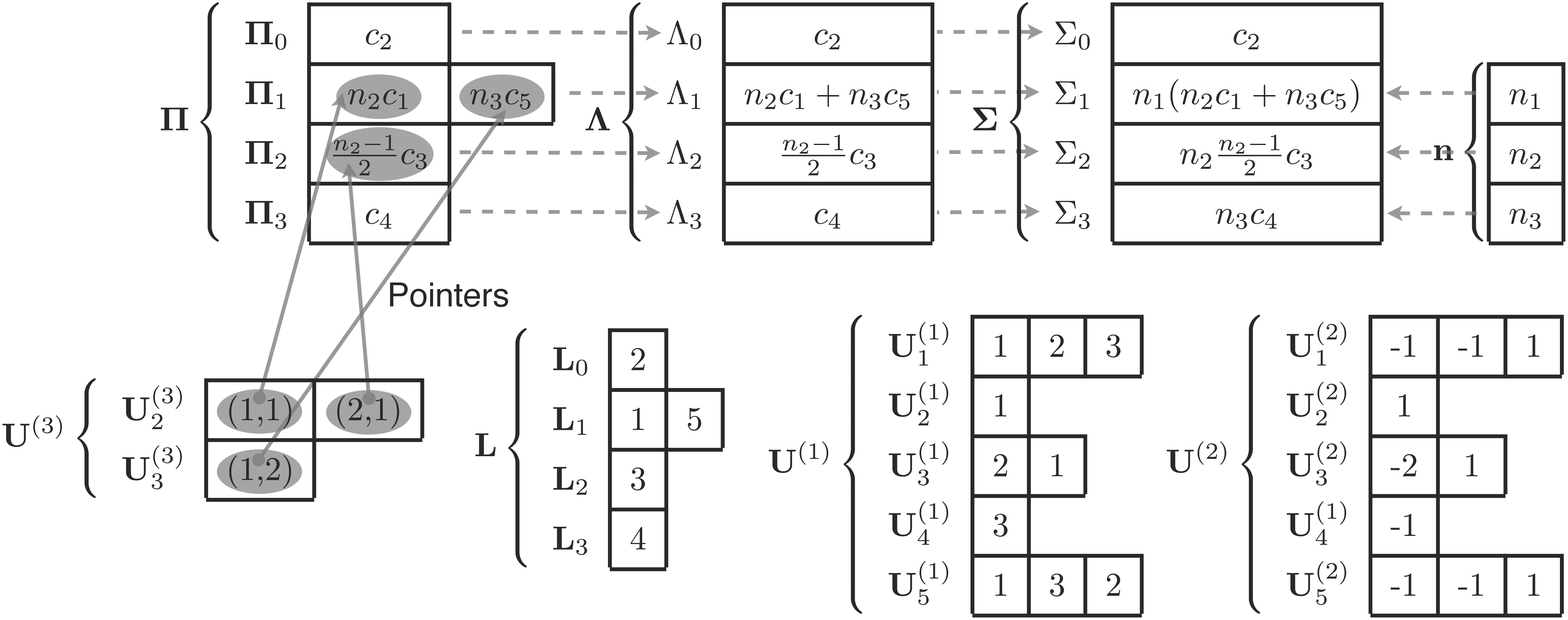, angle=0, height=6cm}}
\subfigure[(b)]{{\label{Example_reac_net}}\epsfig{file=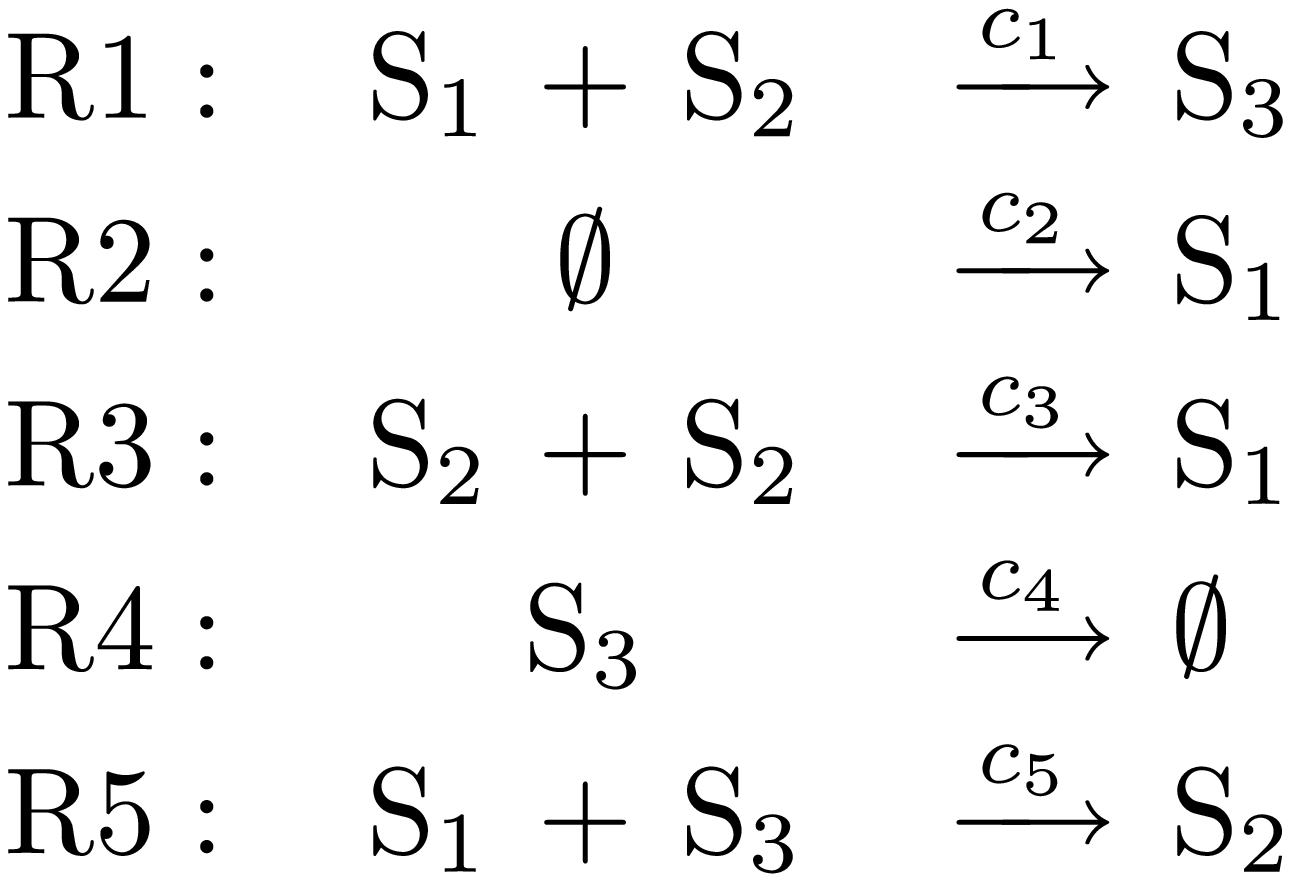, angle=0, height=2.4cm}}
\caption{
(a) Illustration of the data structures in PDM for the example reaction network shown in (b). Note that there may be arrays $\bsym\Pi _i$, $i=1,\ldots,N$, containing at most one negative entry if the corresponding $n_{i}=0$. Indeed, in this example, $\Pi_{2,1}<0$ and $\Lambda_{2}<0$ if $n_{2}=0$. This, however, poses no problem in sampling $I$ and $J$ as all $\Sigma_{i}$ for which $n_i = 0$ are zero and hence the corresponding group indices $I$ are never selected.}
\label{Example_PDM}
\end{center}
\end{figure}

We define the partial propensity of a reaction \rev{with respect to one of its reactants} as the propensity per molecule of \rev{this reactant}. For example, the partial propensity \mbox{$\pi_{\mu}^{(i)}$} of reaction \mbox{$\mu$} with respect to (perhaps the only) reactant S\mbox{$_i$} is \mbox{$a_{\mu}/n_{i}$}, where \mbox{$a_{\mu}$} is the propensity of reaction \mbox{$\mu$} and \mbox{$n_i$} is the number of molecules of S\mbox{$_i$}. The partial propensities of the three elementary reaction types are:
\begin{itemize}
\item Bimolecular reactions (\mbox{$\r{S}_i\,+\,\r{S}_j\,\rightarrow$} Products): \mbox{$a_{\mu}\,=\,n_i\,n_j\,c_{\mu}$} and \mbox{$\pi_{\mu}^{(i)}\,=\,n_j\,c_{\mu}$}, \mbox{$\pi_{\mu}^{(j)}\,=\,n_i\,c_{\mu}$}. \\ 
\rev{If both reactants are of the same species, i.e.~\mbox{$\r{S}_i = \r{S}_j$}, only one partial propensity exists, $\pi_{\mu}^{(i)} \, = \, \frac{1}{2} (n_i - 1)c_{\mu}$ because the reaction degeneracy is \mbox{$\frac{1}{2} n_i (n_i - 1)$}. If $n_i = 0$, the partial propensity becomes negative. As explained in the caption of Fig.~\ref{Example_PDM} this, however, does not require any special treatment.}
\item Unimolecular reactions (\mbox{$\r{S}_i\,\rightarrow$} Products): \mbox{$a_{\mu}\,=\,n_i\,c_{\mu}$} and \mbox{$\pi_{\mu}^{(i)}\,=\,c_{\mu}$}.
\item Source reactions (\mbox{$\emptyset\,\rightarrow$} Products): \mbox{$a_{\mu}\,=\,c_{\mu}$} and \mbox{$\pi_{\mu}^{(0)}\,=\,c_{\mu}$}.
\end{itemize}    
We consider only these elementary reaction types since any reaction with three of more reactants can be treated by decomposing it into a combination of elementary reactions~\cite{Gillespie:1992,Wilhelm:2000,Schneider:2000}. \\   

\subsection{The partial-propensity direct method (PDM)}\label{sec:PDM_description}

In PDM, the index of the next reaction $\mu$ is sampled in a way that is algebraically equivalent to that of DM, as shown in Appendix \ref{pdm:proof}.
The major novelties in PDM are the use of partial propensities and efficient data structures that reduce the number of operations needed to sample $\mu$ and to update the partial propensities. The time to the next reaction is sampled as in DM. We first present the main principles behind the new sampling and update schemes and then describe them in detail. The complete algorithm is given in Table \ref{our_algorithm}.\\

\subsubsection{Main principles behind PDM}

PDM uses partial propensities and groups them in order to efficiently sample the index of the next reaction and update the partial propensities after a reaction has fired. For the sampling step, the partial propensities are grouped according to the index of the factored-out reactant, yielding at most \mbox{$N+1$} groups of size $O(N)$. Sampling then proceeds in two steps: we first sample the index of the group before sampling the actual partial propensity inside that group. This grouping scheme reduces the number of operations needed for sampling the next reaction using a concept that is reminiscent of two-dimensional cell lists \cite{Hockney:1988}. If all partial propensities are in the same group, or if every group contains only a single partial propensity, the sampling step of PDM is no more efficient than that of DM. These cases, however, can only occur if the function \mbox{$M(N)$} is \mbox{$O(N)$} (see for example the linear chain model) and both PDM and DM hence have a computational cost of $O(N)$ for sampling the index of the next reaction. \\

After the selected reaction has been executed, we use a dependency graph over species (partial propensities), rather than reactions, to find all partial propensities that need to be updated. This is possible because partial propensities depend on the population of at most one species, and is analogous to a Verlet list \cite{Verlet:1967}. This limits the number of updates to be $O(N)$. 
In addition, partial propensities of unimolecular reactions are constant and never need to be updated. 
In weakly coupled networks, where the degree of coupling is $O(1)$, the scaling of the computational cost of the update becomes equal to that of methods that use dependency graphs over reactions, such as SSA-CR, ODM, and SDM. \\

We illustrate the sampling scheme of PDM in a simple protein aggregation example. Consider proteins that aggregate to form at most tetrameric complexes. There are \mbox{$N=4$} species in the reaction network: monomers, dimers, trimers, and tetramers. All species except tetramers can aggregate in all possible combinations to form multimeric complexes (4 bimolecular reactions). In addition, all multimeric complexes can dissociate into any possible combination of two smaller units (4 unimolecular reactions) and monomers are constantly produced (1 source reaction). This reaction network is described by $M=9$ partial propensities $(\pi_{1}^{(0)})$, $(\pi_{2}^{(1)}$, $\pi_{3}^{(1)}$, $\pi_{4}^{(1)})$, $(\pi_{5}^{(2)}$, $\pi_{6}^{(2)})$, $(\pi_{7}^{(3)})$, $(\pi_{8}^{(4)}$, $\pi_{9}^{(4)})$. Grouping the partial propensities according to the index of the factored-out reactant given in the superscript, we obtain 5 (\mbox{$=N+1$}) groups as indicated by the parentheses. Along with each group, we store the sum of all partial propensities inside it. Using a random number, we sample the group that contains the next reaction, before finding the corresponding partial propensity inside that group. Assume that in our example reaction 7 is to fire next. The search depth to find the group index is 4 and we need 1 additional operation to find the partial propensity $(\pi_{7}^{(3)})$. PDM thus requires 5 operations to sample the next reaction in this network of 9 reactions. The average search depth of sampling the next reaction in this example is \mbox{$37/9\approx 4.1$}.  \\      

In the next section, we formally describe PDM and its data structures.     

\subsubsection{Detailed description of the PDM algorithm}

All partial propensities are stored in the ``partial-propensity structure'' \mbox{${\bsym\Pi}=\left\{ \bsym{\Pi}_i \right\}_{i=0}^N$} as a one-dimensional array of one-dimensional arrays \mbox{${\bsym\Pi}_i$}. Each array \mbox{${\bsym\Pi}_i$} contains the partial propensities belonging to group $i$. 
The partial propensities of source reactions are stored as consecutive entries of the 0\mbox{$^\text{th}$} array \mbox{$\boldsymbol \Pi _0$}. The partial propensities of all reactions that have species S\mbox{$_1$} as one of its reactants are stored as consecutive entries of \mbox{$\boldsymbol \Pi _1$}. In general, the \mbox{$i^\text{th}$} array \mbox{$\boldsymbol \Pi_{i}$} contains the partial propensities of all reactions that have S\mbox{$_i$} as a reactant, provided these reactions have not yet been included in any of the previous \mbox{$\boldsymbol \Pi_{j<i}$}. That is, out of the two partial propensities of a reaction \mbox{$\mu$} with S\mbox{$_i$} and S\mbox{$_j$} as its reactants, \mbox{$\pi_{\mu}^{(i)}$} is part of \mbox{$\boldsymbol \Pi_{i}$} if \mbox{$i<j$}, and \mbox{$\pi_{\mu}^{(j)}$} is not stored anywhere. Notice that, since the different \mbox{$\boldsymbol \Pi_{i}$}'s can be of different length, storing them as an array of arrays is more (memory) efficient than using a matrix (i.e.~a two-dimensional array). The reaction indices of the partial propensities in ${\boldsymbol \Pi}$ are stored in a look-up table \mbox{$\textbf{L}=\{\mathrm{\mathbf{L}}_i\}_{i=0}^{N}$}, which is also an array of arrays. This makes every reaction \rev{\mbox{$\mu$}} identifiable by a unique pair of indices, \rev{a group index} $I$ and \rev{an element index} $J$, such that the partial propensity of reaction \mbox{$\mu =\r{L}_{I,J}$} is stored in \mbox{$\Pi_{I,J}$}. \\

We further define the ``group-sum array'' $\boldsymbol \Lambda$, storing the sums of the partial propensities in each group \mbox{$\boldsymbol \Pi_{i}$}, thus \mbox{$\Lambda _i = \sum_{j} \Pi_{i,j}$}, \mbox{$i\,=\,0,\ldots, N$}. In addition, we also define \mbox{$\bsym{\Sigma}$}, the array of the total propensities of all groups, as \mbox{$\Sigma_i = n_{i}\Lambda _i$}, \mbox{$i\,=\,1,\ldots, N$}, and \mbox{$\Sigma _0 = \Lambda _0$}. The total propensity of all reactions is then \mbox{$a=\sum_{i=0}^{N}\Sigma_i$}. The use of \mbox{$\bsym\Lambda$} avoids having to recompute the sum of all partial propensities in \mbox{$\boldsymbol \Pi_{i}$} after one of them has changed. Rather, the same change is also applied to \mbox{$\Lambda _i$} and computing the new \mbox{$\Sigma _i$} only requires a single multiplication by \mbox{$n_i$}. Using these data structures and a single uniformly distributed random number \rev{\mbox{$r_1\,\in\,[0,1)$}}, the next reaction $\mu$ can efficiently be sampled in two steps: (1) sampling the \rev{group} index $I$ such that

\be
\label{row_mu}
I ={\operatorname{min}}\left[ I' \,\, : \,\, r_{1}a < \sum_{i=0}^{I'} \Sigma_{i} \right]
\,
\ee
and (2) sampling \rev{the} \rev{element} index $J$ in ${\boldsymbol \Pi}_I$ such that
\be
J ={\operatorname{min}}\left[ 
J ' \,\, : \,\, r_{1}a < \sum_{j=1}^{J'}n_{I}\Pi_{I,j} + \left( \sum_{i=0}^{I} \Sigma_{i}\right) - \Sigma_{I}  \right] \,.
\label{col_mu}
\ee
(See Appendix~\ref{pdm:proof} for a proof of the equivalence of this sampling scheme to that of DM.) Using the temporary variables
\be
\label{col_mu1}
\Phi = \sum_{i=0}^{I} \Sigma_{i},\qquad
\Psi = \frac{r_1 a - \Phi + \Sigma _I}{n_I} \, ,
\ee
Eq.~\ref{col_mu} can be efficiently implemented as
\be
\label{col_mu3}
J ={\operatorname{min}}\left[ 
J' \,\, : \,\, \Psi < \sum_{j=1}^{J'}\Pi_{I,j} \right] \, .
\ee

The indices $I$ and $J$ are then translated back to the reaction index $\mu$ using the look-up table $\mathbf{L}$, thus \mbox{$\mu = \r{L}_{I,J}$}. \\

Once a reaction has been executed, $\vec{n}$, $\boldsymbol \Pi$, $\boldsymbol \Lambda$, and $\boldsymbol \Sigma$ need to be updated. This is efficiently done using three update structures:
\begin{enumerate}
\item[\mbox{$\mathrm{\mathbf{U}^{(1)}}$}] is a array of $M$ arrays, where the \mbox{$i^\text{th}$} array contains the indices of all species involved in the \mbox{$i^\text{th}$} reaction. 
\item[\mbox{$\mathrm{\mathbf{U}^{(2)}}$}] is a array of $M$ arrays containing the corresponding stoichiometry (the change in population of each species upon reaction) of the species stored in \mbox{$\mathrm{\mathbf{U}^{(1)}}$}.
\item[\mbox{$\mathrm{\mathbf{U}^{(3)}}$}] is a array of $N$ arrays, where the \mbox{$i^\text{th}$} array contains the indices of all entries in $\boldsymbol \Pi$ that depend on \mbox{$n_{i}$}, thus: 
\end{enumerate}
\bea
\mathbf{U}^{(3)}\,=\,
\left\{\begin{array}{l}
\mathbf{U}_1^{(3)} \,=\,\left(i^{1}_1,j^{1}_1 \quad i^{1}_2,j^{1}_2 \quad\ldots\quad\ldots\quad\ldots\,\right)\\
\mathbf{U}_2^{(3)} \,=\,\left(i^{2}_1,j^{2}_1 \quad i^{2}_2,j^{2}_2 \quad\ldots\,\right)\\
\vdots\\
\mathbf{U}_N^{(3)} \,=\,\left(i^{N}_1,j^{N}_1 \quad i^{N}_2,j^{N}_2 \quad\ldots\quad\ldots\,\right).
\end{array}\right.
\label{U3}
\eea

When a reaction is executed, the populations of the species involved in this reaction change. Hence, all entries in $\boldsymbol \Pi$ that depend on these populations need to be updated. After each reaction, we use \mbox{$\mathrm{\mathbf{U}}^{(1)}$} to determine the indices of all species involved in this reaction. The stoichiometry is then looked up in \mbox{$\mathrm{\mathbf{U}}^{(2)}$} and the population $\vec{n}$ is updated. Subsequently, \mbox{$\mathrm{\mathbf{U}^{(3)}}$} is used to locate the affected entries in $\boldsymbol \Pi$ and recompute them. The two data structures \mbox{$\mathrm{\mathbf{U}}^{(1)}$} and \mbox{$\mathrm{\mathbf{U}}^{(2)}$} are a sparse representation of the stoichiometry matrix, and \mbox{$\mathrm{\mathbf{U}}^{(3)}$} represents the dependency graph over species. Since the partial propensities of unimolecular and source reactions are constant and need never be updated, \mbox{$\mathrm{\mathbf{U}^{(3)}}$} only contains the indices of the partial propensities of bimolecular reactions.
The size of \mbox{$\mathbf{U}^{(3)}$} is at most a factor of $N$ smaller than that of the corresponding dependency graph over reactions, since partial propensities depend on the population of at most one species.  
Figure \ref{Example_PDM} summarizes the data structures used in PDM for an example reaction network. The complete algorithm is given in Table~\ref{our_algorithm}. 
Overall, PDM's computational cost is \mbox{$O(N)$} and its memory requirement is \mbox{$O(M)$}, irrespective of the degree of coupling (see Appendix \ref{app:complexity}).  

\subsection{The sorting partial-propensity direct method (SPDM)}

The sorting partial-propensity direct method (SPDM) is the partial-propensity variant of SDM~\cite{McCollum:2006}. In SPDM, the group and element indices $I$ and $J$ are bubbled up whenever the reaction \mbox{$\mu = \r{L}_{I,J}$} fires. The reordered indices are stored in an array for $I$, and an array of arrays of the size of $\boldsymbol \Pi$ for the $J$'s. This requires an additional \mbox{$N+M$} memory, but further reduces the search depth to sample the next reaction, especially in a multiscale (stiff) network. The computational cost of SPDM is also $O(N)$ (see Appendix \ref{app:complexity}), but with a possibly reduced pre-factor. 

\section{Benchmarks}\label{sec:benchmarks}

We benchmark the computational performance of PDM and SPDM using four chemical reaction networks that are prototypical of: (a) strongly coupled reaction networks, (b) strongly coupled reaction networks comprising only bimolecular reactions, (c) weakly coupled reaction networks, and (d) multiscale biological networks. The first two benchmarks consider strongly coupled networks where the degree of coupling scales with system size (see column ``degree of coupling'' in Table \ref{table_test_cases}). The first benchmark consists of a colloidal aggregation model. The second benchmark considers a network of only bimolecular reactions, where none of the partial propensities are constant. In the third benchmark, we compare PDM and SPDM to SDM on the linear chain model, a weakly coupled reaction network with the minimal degree of coupling, for which SDM was reported to be very efficient \cite{Cao:2004,McCollum:2006}. The fourth benchmark considers the heat-shock response model, a small multiscale (stiff) biological reaction network of fixed size. The benchmark problems are defined in detail in Appendix \ref{app:testproblems}, where also the respective partial-propensity structures $\mathbf{\Pi}$ are given. \\

All tested SSA formulations are implemented in C++ using the random-number generator of the GSL library and compiled using the GNU C++ compiler version 4.0.1 with the O3 optimization flag. All timings are determined using a nanosecond-resolution timer (the {\tt{mach\_absolute\_time()}} system call) on a MacOS X 10.4.11 workstation with a 3\,GHz dual-core Intel Xeon processor, 8\,GB of memory, and a 4\,MB L2 cache. For each test case, we report both the memory requirement and the average CPU time per reaction (i.e.~per time step), $\Theta$. $\Theta$ is defined as the CPU time (identical to wall-clock time in our case) needed to simulate the system up to final time $T$, divided by the total number of reactions executed during the simulation, and averaged over independent runs. The time $\Theta$ does not include the initialization of the data structures (step 1 in Table \ref{our_algorithm}) as this is done only once and is not part of the time loop. \\

We explain the benchmark results in terms of the computational cost of the individual steps of the algorithms. We distinguish three steps: (a) sampling the index of the next reaction, (b) updating the population, and (c) updating the partial propensities (for PDM and SPDM) or the propensities (for SDM). The computational costs of these steps are quantified separately and the overall timings are then explained as a weighted sum of: 
\begin{itemize}
\item \mbox{$\mathcal{C}_\mu$}: The number of operations required to sample the index of the next reaction (for PDM, this is step 2 in Table~\ref{our_algorithm}). 
\item \mbox{$\mathcal{C}_\mathbf{n}$}: The number of elements of the population $\vec{n}$ that need to be updated after executing a reaction (for PDM, this is step 4 in Table~\ref{our_algorithm}). 
\item \mbox{$\mathcal{C}_\r{P}$}: The number of (partial) propensities that need to be updated after executing a reaction (for PDM, this is step 5.2.2 in Table~\ref{our_algorithm}). 
\end{itemize}   
The expressions for these elementary costs are given in Table \ref{table_Cs} as determined by independently fitting models for the scaling of the algorithms to the measured operation counts, averaged over 100 independent runs of each test problem. In all cases, the models used for the computational cost explain the data with a correlation coefficient of at least 0.98. The benchmark results are then explained by fitting the weights of the cost superposition \mbox{$a \mathcal{C}_\mu + b \mathcal{C}_\mathbf{n} + c \mathcal{C}_\r{P}$} to the measured scaling curves $\Theta(N)$ using the expressions given in Table \ref{table_Cs}. In order to preserve the relative weights of the data points, all fits are done on a linear scale, even though the results are plotted on a logarithmic scale for two of the benchmarks. All these fits also have a correlation coefficient of at least 0.98. Explaining the timing results as a superposition of elementary costs allows determining which part of an algorithm is responsible for a particular speedup or scaling behavior, and what the relative contributions of the three algorithmic steps are to the overall computational cost. \\ 

The memory requirements of the algorithms are reported in Table \ref{table_Ms} for all benchmark cases. These numbers were derived analytically from the size of the individual data structures. 

\subsection{Strongly coupled reaction network: colloidal aggregation model}

We use the colloidal aggregation model in Appendix \ref{app:testproblems}\ref{app:aggregation}~\cite{Meakin:1988,Lin:1989,Lin:1990a,Axford:1996,Turner:2005} as a first example of a strongly coupled reaction network. This reaction network can be used to model, e.g., colloidal aggregation of solvated proteins, nano-beads, or viruses. For $N$ chemical species it consists of \mbox{$M = \lfloor{\frac{N^2}{2}}\rfloor$} reactions and the maximum out-degree of the dependency graph is \mbox{$3N-7$} and hence scales with system size (see Table \ref{table_test_cases}).  \\ 

The colloidal aggregation model is simulated up to time \mbox{$T = 100$} with specific probability rates \mbox{$c_{n,m} = 1$} and \mbox{$\bar{c}_{p,q} = 1$}. At time \mbox{$t = 0$}, \mbox{$n_i = N\delta_{1,i}$}. 
The scaling of $\Theta$ for PDM, SPDM, and SDM with system size is shown in Fig.~\ref{Aggregation_model1}(a), averaged over 100 independent runs. \mbox{$\Theta^{\r{PDM}}$} and \mbox{$\Theta^{\r{SPDM}}$} are \mbox{$O(N^{0.5})$} for small $N$ (less than about 100) and \mbox{$O(N)$} for large $N$. \mbox{$\Theta^{\r{SDM}}$} is \mbox{$O(N^{2})$}. The pre-factor of \mbox{$\Theta^{\r{SPDM}}$} is similar to that of \mbox{$\Theta^{\r{PDM}}$}, since in this network \mbox{$\mathcal{C}_\mu$} is not significantly reduced by the dynamic sorting (Table \ref{table_Cs}). The memory requirements of PDM and SPDM are \mbox{$O(N^2)= O(M)$}, that of SDM is \mbox{$O(N^3)= O(NM)$} (Table \ref{table_Ms}). \\  

In summary, the computational costs of both PDM and SPDM are $O(N)$. This scaling is mediated by all three cost components. The use of partial propensities renders the scaling of the sampling cost \mbox{$\mathcal{C}_\mu$} \mbox{$O(N)$} (see Table \ref{table_Cs}). The cost \mbox{$\mathcal{C}_\r{P}$} for updating the partial propensities is \mbox{$O(N^{0.5})$} (Table \ref{table_Cs}), since the use of partial propensities allows formulating a dependency graph over species, rather than reactions, and unimolecular reactions have constant partial propensities. This leads to a smaller number of updates needed as shown in Fig.~\ref{Aggregation_updates}(a). 

\subsection{Strongly coupled network of bimolecular reactions}

The network in Appendix \ref{app:testproblems}\ref{app:tightly} consists of \mbox{$M=\frac{N}{2}(N-1)$} strongly coupled bimolecular reactions, such that none of the partial propensities are constant. Both the minimum and the maximum out-degrees of the dependency graph in this case are \mbox{$4N-10$}, scaling faster with $N$ than in the previous case (see Table \ref{table_test_cases}). \\   

We simulate this network up to time \mbox{$T = 0.001$} with all specific probability rates \mbox{$c_i = 1$}. At \mbox{$t = 0$}, \mbox{$n_i = 100 (\delta_{N-4,i} + \delta_{N-3,i} + \delta_{N-2,i} + \delta_{N-1,i} + \delta_{N,i})$}. The scaling of $\Theta$ for PDM, SPDM, and SDM with system size is shown in Fig.~\ref{Tightly_coupled_model}(b), averaged over 100 independent runs. \mbox{$\Theta^{\r{PDM}}$} and \mbox{$\Theta^{\r{SPDM}}$} are \mbox{$O(N)$}, whereas \mbox{$\Theta^{\r{SDM}}$} is \mbox{$O(N^2)$}. The pre-factors of PDM and SPDM are comparable. The memory requirements of PDM and SPDM are \mbox{$O(N^2)= O(M)$}, that of SDM is \mbox{$O(N^3)= O(NM)$} (see Table \ref{table_Ms}). \\

In summary, the computational costs of PDM and SPDM are $O(N)$ for this strongly coupled, purely bimolecular network. The scaling is again mediated by all three cost components. Grouping the partial propensities renders the sampling cost \mbox{$\mathcal{C}_\mu$} \mbox{$O(N)$} (see Table \ref{table_Cs}). Because none of the partial propensities are constant, the update costs \mbox{$\mathcal{C}_\r{P}$} of PDM and SPDM are $O(N)$, as in SDM, albeit with a pre-factor that is \mbox{$\approx 2.5$} times smaller than that in SDM. One reason for this smaller pre-factor is the smaller number of updates needed upon reactions firing, as shown in Fig.~\ref{Tightly_coupled_updates}(b). This is due to the fact that partial propensities of bimolecular reactions depend on the population of only one species, which reduces the number of combinations that need to be updated.

\subsection{Weakly coupled reaction network: linear chain model}\label{sec:linchain}

We benchmark PDM and SPDM on a weakly coupled model in order to assess their limitations in cases where other SSA formulations might be more efficient. We choose the linear chain model defined in Appendix \ref{app:testproblems}\ref{app:linearchain} since it is the most weakly coupled reaction network possible and it has been used as a model for isolated signal transduction networks~\cite{Albert:2005}. 
For $M$ reactions, it involves the minimum number of species \mbox{$N=M+1$}, and the maximum out-degree of the dependency graph is constant at the minimum possible value of 2 (see Table \ref{table_test_cases}), since every reaction at most influences the population of its only reactant and of the only reactant of the subsequent reaction. \\

We simulate the linear chain model to a final time of \mbox{$T = 1000$} with all specific probability rates \mbox{$c_i = 1$}. At time \mbox{$t = 0$}, \mbox{$n_i = 10000 \delta_{1,i}$}. Figure~\ref{Linear_model1}(c) presents the scaling of the CPU time with system size for PDM, SPDM, and SDM, averaged over 100 independent runs. \mbox{$\Theta^{\r{PDM}}$} scales linearly with $N$ and \mbox{$\Theta^{\r{SPDM}}$} with \mbox{$N^{0.5}$}. \mbox{$\Theta^{\r{SDM}}$} is \mbox{$O(N)$} with a pre-factor that is more than 4 times larger than that of \mbox{$\Theta^{\r{PDM}}$}. This difference in pre-factor is mainly caused by PDM having smaller \mbox{$\mathcal{C}_\mathbf{n}$} and \mbox{$\mathcal{C}_\r{P}$} (Table \ref{table_Cs}). \mbox{$\mathcal{C}_\mu$}, however, scales worse for PDM than for SDM due to the dynamic sorting in SDM. This is overcome in SPDM, where  \mbox{$\mathcal{C}_\mu$} is \mbox{$O(N^{0.5})$}, as in SDM. The memory requirements of SPDM and PDM are \mbox{$O(N)= O(M)$}, that of SDM is \mbox{$O(N^2)= O(NM)$} (Table \ref{table_Ms}). \\

In summary, the computational costs of PDM and SPDM on the weakly coupled linear chain model are governed by (a) updating the population $\vec{n}$ using a sparse stoichiometry representation and (b) never needing to update the partial propensities of unimolecular reactions. Since the linear chain model contains only unimolecular reactions, none of the partial propensities ever needs to be updated, leading to an update cost of \mbox{$\mathcal{C}_\r{P} = 0$} (see Table \ref{table_Cs}). While we have implemented SDM according to the original publication \cite{McCollum:2006}, we note that if one uses a sparse representation of the stoichiometry matrix also in SDM, point (a) vanishes and \mbox{$\mathcal{C}_\mathbf{n}=2$} also for SDM. A sparse-stoichiometry SDM would thus have the same scaling of the computational cost on the linear chain model as would SPDM, outperforming PDM.      

\subsection{Multi-scale biological network: heat-shock response in \textit{Escherichia coli}}

We assess the performance of PDM and SPDM on a small, fixed-size multiscale reaction network. We choose the heat-shock response model since it has also been used to benchmark previous methods, including ODM~\cite{Cao:2004} and SDM~\cite{McCollum:2006}. The model describes one of the mechanisms used by the bacterium {\textit{E.~coli}} to protect itself against a variety of environmental stresses that are potentially harmful to the structural integrity of its proteins. The heat-shock response (HSR) system reacts to this by rapidly synthesizing heat-shock proteins. The heat-shock sigma factor protein \mbox{$\sigma^{32}$} activates the HSR by inducing the transcription of heat-shock genes. The heat-shock response model is a small multiscale reaction network (the specific probability rates span 8 orders of magnitude) with \mbox{$N=28$} chemical species, \mbox{$M=61$} reactions, and a maximum out-degree of the dependency graph of 11 (see Table \ref{table_test_cases}). For a detailed description of the model, we refer to Kurata et al.~\cite{Kurata:2001}\\          

\begin{figure}[htbp]
\begin{center}
\subfigure[(a) Colloidal aggregation model]{{\label{Aggregation_updates}}\epsfig{file=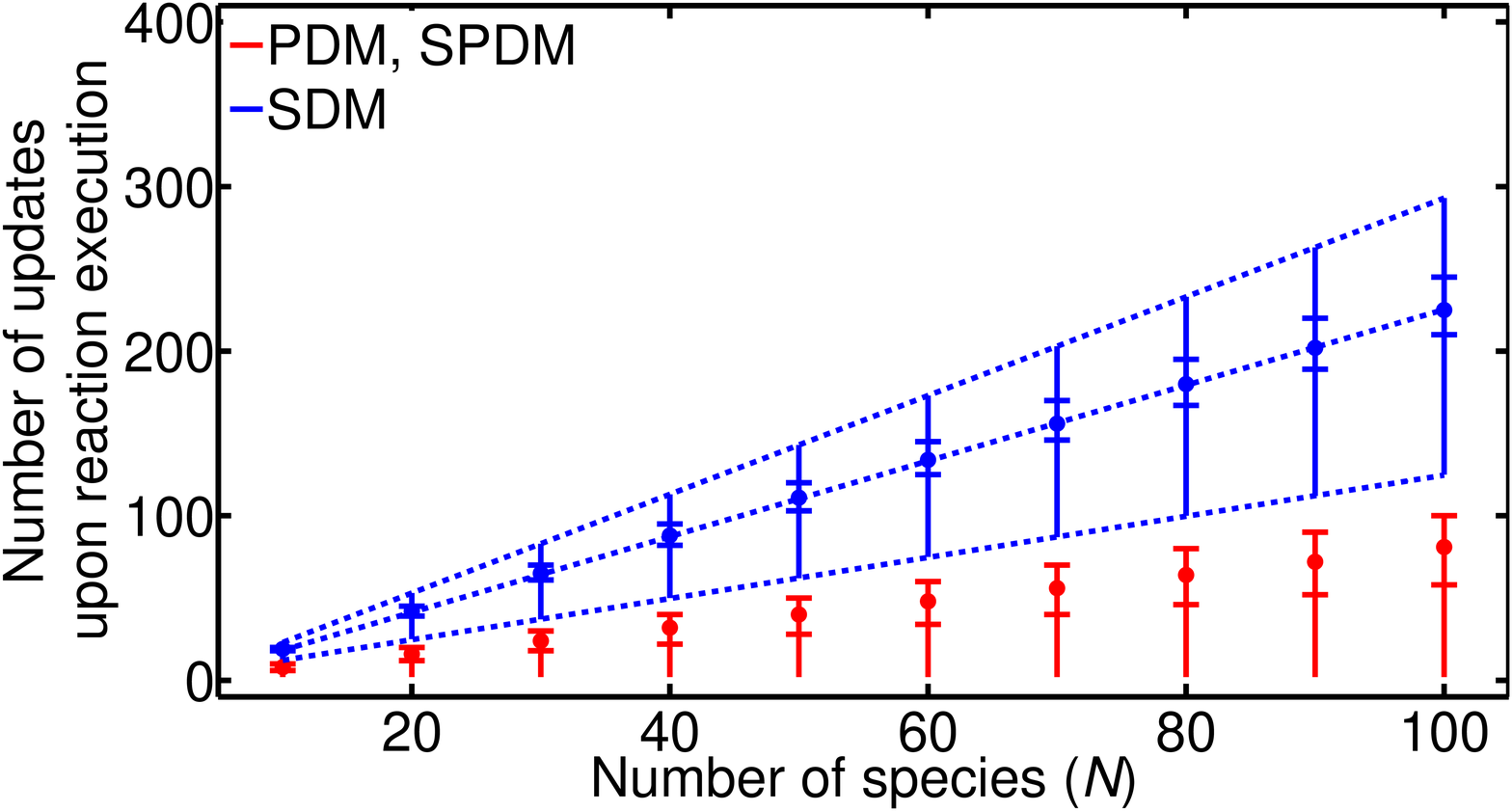, angle=0, width=8.3cm}}
\subfigure[(b) Network of bimolecular reactions]{{\label{Tightly_coupled_updates}}\epsfig{file=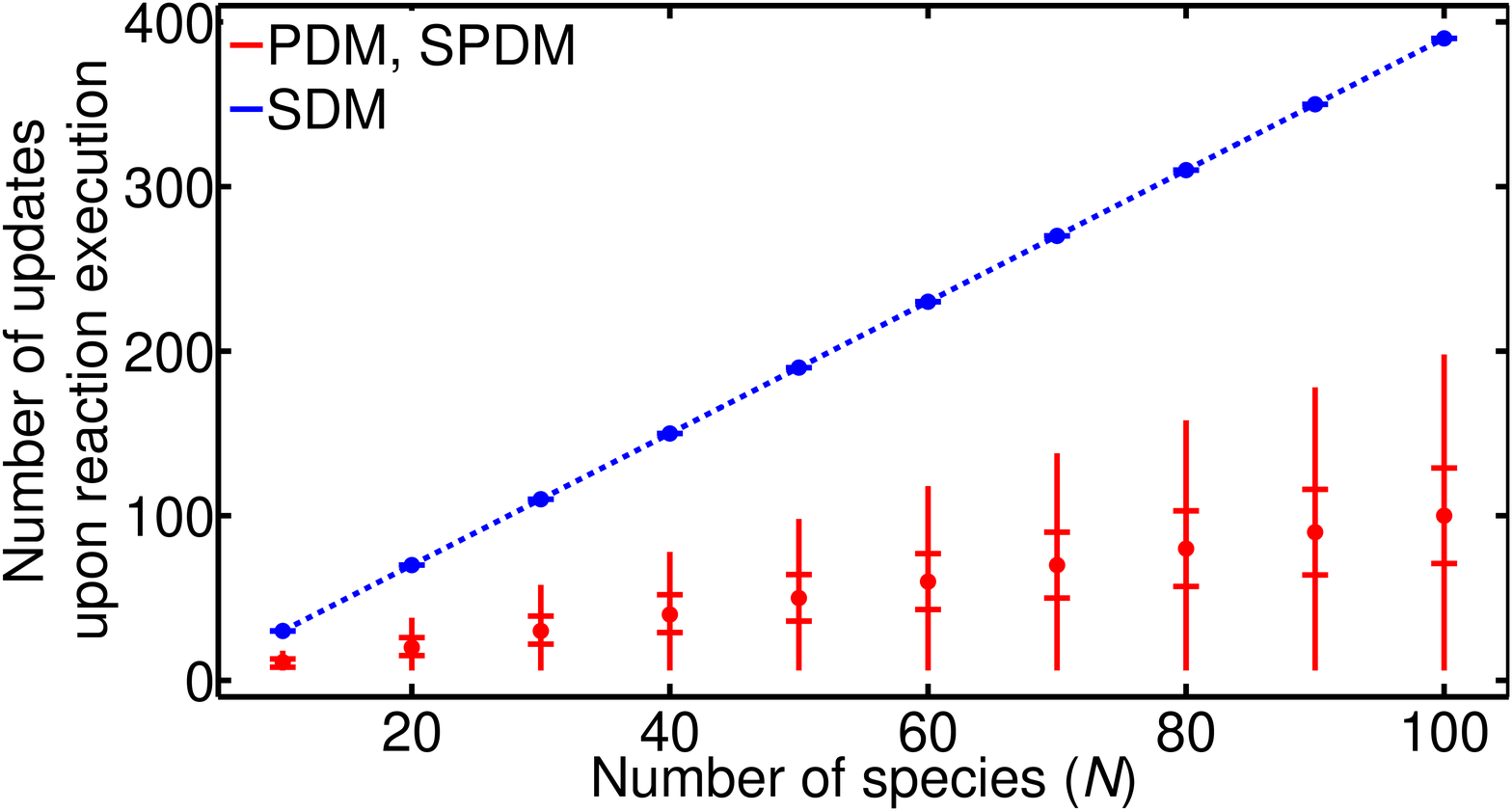, angle=0, width=8.3cm}}
\subfigure[(c) Heat shock response model]{\label{HSR_updates}\epsfig{file=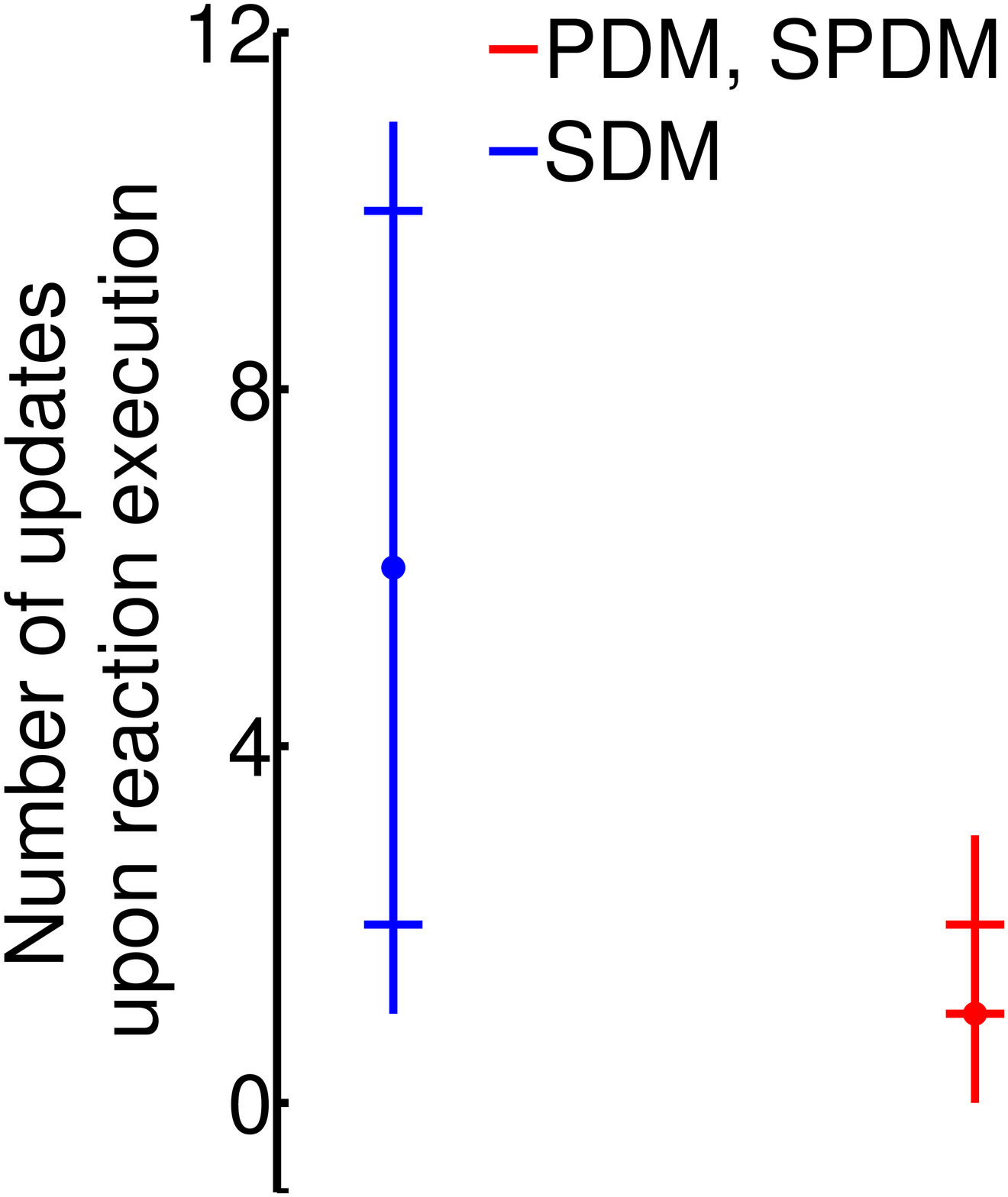, angle=0, width=4.5cm}}
\caption{Measured distributions of the number of partial propensities (for PDM and SPDM, red line) and propensities (for SDM, blue line) that need to be updated after firing any reaction of: (a) the colloidal aggregation model, (b) the network of bimolecular reactions, and (c) the heat-shock response model. Dots indicate medians, horizontal bars the upper and lower quartiles, and vertical bars the upper and lower extrema (maximum and minimum). The dotted lines denote the minimum, average and maximum degree of coupling $k$ of the reaction networks (see Table \ref{table_test_cases}). 
The number of updates in SDM \cite{McCollum:2006} using a dependency graph is governed by the degree of coupling of the network. In PDM and SPDM, less updates need to be performed since partial propensities depend on the population of at most one species and are constant for unimolecular reactions.}
\label{Upsilon_vs_N_plot}\label{Aggregation_updates}\label{Tightly_coupled_updates}\label{HSR_updates}
\end{center}
\end{figure}

We simulate the HSR model for \mbox{$T = 500$} seconds. During this time, approximately 46 million reactions are executed. 
For a single run, we measure \mbox{$\Theta^{\r{PDM}}=0.256\,\mu$}s and \mbox{$\Theta^{\r{SDM}}=0.272\,\mu$}s. This corresponds to a simulated 3.68 million reactions per second of CPU time for SDM and 3.89 million reactions per second for PDM. Hence, PDM is about 6\% faster than SDM. This speed-up is mainly due to  a smaller \mbox{$\mathcal{C}_\r{P}$} in PDM (see Fig.~\ref{Upsilon_vs_N_plot}(c) for the distribution of updates over all reactions) since the partial propensities of unimolecular reactions never need to be updated. The speed-up is, however, modest because \mbox{$\mathcal{C}_\mu$} of PDM is \mbox{$\approx 4.6$} times larger than that of SDM (Table \ref{table_Cs}). This is due to the fact that 95\% of all reaction firings are caused by a small subset of only 6 reactions. This multiscale network thus strongly benefits from the dynamic sorting used in SDM. This advantage can be recovered in SPDM, where \mbox{$\mathcal{C}_\mu$} is comparable to that of SDM, and \mbox{$\Theta^{\r{SPDM}}=0.245\,\mu$}s (4.08 million reactions per second). This makes SPDM 11\% faster than SDM on this small network.      

\section{Conclusions and Discussion}\label{sec:conclusion}

The stochastic simulation algorithm (SSA)~\cite{Gillespie:1976,Gillespie:1977,Gillespie:1992} is widely used for computational stochastic reaction kinetics in chemistry, physics, biology, and systems biology. It is included in most existing stochastic simulation software packages and is standard in courses on computational chemical kinetics. Due to this importance, several variants of the original SSA formulation have been published that reduce the computational costs of the sampling and update steps. When simulating weakly coupled reaction networks, where the maximum number of reactions that are influenced by any reaction is constant with system size, the computational cost of the sampling step has been reduced to be $O(\log _2 M)$ \cite{Gibson:2000}, where $M$ is the total number of reactions, and even to $O(1)$ under some conditions for the propensity distribution \cite{Slepoy:2008}. Using dependency graphs, also the update step has been reduced to be $O(1)$ for weakly coupled networks \cite{Cao:2004,McCollum:2006,Slepoy:2008}.  
For strongly coupled reaction networks, where the degree of coupling increases with system size and can be as large as the total number of reactions, all previous exact SSA formulations have a computational cost that is $O(M)$. \\

We have introduced a new quantity called {\em partial propensity} and have used it to construct two novel formulations of the exact SSA: PDM and its sorting variant SPDM. Both are algebraically equivalent to DM and yield the same population trajectories $\vec{n}(t)$ as to those produced by DM. In our formulation of partial propensities, we have limited ourselves to elementary chemical reactions. Since their partial propensities depend on the population of at most one species, both new SSA formulations have a computational cost that scales at most linearly with the number of species rather than the number of reactions, independently of the degree of coupling. This is particularly advantageous in strongly coupled reaction networks, where the number of reactions $M$ grows faster than the number of species $N$ with system size. On networks of fixed size, PDM and SPDM are especially efficient when \mbox{$M\gg N$}.
PDM's computational cost is $O(N)$, which is made possible by appropriately grouping the partial propensities in the sampling step and formulating a dependency graph over species rather than reactions in the update step. Moreover, the partial propensities of unimolecular reactions and source reactions are constant and never need to be updated. This further reduces the size of the dependency graph and the computational cost of the update step. To our knowledge, PDM is the first SSA formulation that has a computational cost that is $O(N)$, irrespective of the degree of coupling of the reaction network. In the case of multiscale networks, the computational cost of SPDM is smaller than that of PDM. \\

We have benchmarked PDM and SPDM on four test cases of various degrees of coupling. The first two benchmarks considered strongly coupled networks, where the degree of coupling scales proportionally to the number of species. The third benchmark considered the most weakly coupled network possible, where several other SSA formulations might be more efficient. Finally, the fourth benchmark considered a small biological multiscale network. 
These benchmarks allowed estimating the scaling of the computational cost with system size and the cost contributions from reaction sampling, population update, and partial-propensity update. The results showed that (a) the overall computational costs of PDM and SPDM are $O(N)$, even for strongly coupled networks, (b) on very weakly coupled networks, SPDM is competitive compared to SDM, (c) on multiscale networks SPDM outperforms PDM, and (d) the memory requirements of PDM and SPDM are $O(M)$ in all cases, and hence not larger than those of any other exact SSA formulations. \\

Currently, PDM and SPDM have a number of limitations. The most important limitation is that the presented formulation of partial propensities is only applicable to elementary chemical reactions. Any higher-order chemical reaction can always be broken down into elementary reactions at the expense of increasing system size. In applications such as 
population ecology or social science, the idea of partial propensities can, however, only be used if the (generalized) reactions are at most binary and one species can be factored out, i.e.~if the propensity for every reaction between species $\text{S}_i$ and $\text{S}_j$ can be written as $a_\mu = c_\mu n_i \tilde{h}(n_j)$. Besides this structural limitation, the computational performance of the particular algorithms presented here can be limited in several situations. One of them is the simulation of very small networks, where the overhead of the data structures involved in PDM and SPDM may not be amortized by the gain in efficiency and a simulation using DM may be more efficient. In multiscale networks, where the propensities span several orders of magnitude, PDM is slower than SPDM. In multiscale networks where a small subset ($\ll N$) of all reactions accounts for almost all of the reaction firings, however, the overhead of the data structures involved in SPDM, including their initialization, may not be amortized by the gain in efficiency.
Finally, PDM and SPDM were designed to have a computational cost that scales linearly with the number of species rather than the number of reactions. In reaction networks in which the number of reactions grows sub-linearly with the number of species, this becomes a disadvantage. In such cases, SSA formulations that scale with the number of reactions are favorable.  \\

The classification of reaction networks according to their ``difficulty'' is still largely an open question. Besides system size, degree of coupling, and multiscaling  (spectrum of time scales), there might also be other network properties that influence the computational cost of the various SSA formulations. Automatized selection of the most efficient SSA formulation for a given network would require both a systematic classification of networks and a prediction of the computational cost of SSA formulations based on network properties. 
This might require a more detailed cost analysis of the algorithms and a set of standard benchmark problems that are designed to cover the entire range of performance-relevant parameters. \\

Taken together, our results suggest that PDM and SPDM can potentially offer significant performance improvements especially in strongly coupled networks, including the simulation of colloidal aggregation~\cite{Meakin:1988,Lin:1989,Lin:1990a,Axford:1996,Turner:2005}, Becker-D\"oring-like nucleation-and-growth reactions~\cite{Wattis:2009}, and scale-free biochemical reaction networks, where certain hubs are strongly coupled \cite{Jeong:2000,Strogatz:2001,Albert:2002,Albert:2005}. Finally, the use of partial propensities is not limited to exact SSA formulations, and we also expect approximate methods to benefit from it. 
The software implementations of PDM and SPDM will be made available as open source on the web page of the authors. 

\section{Acknowledgments}
We thank Dr.~Hong Li and Prof.~Dr.~Linda Petzold, University of California at Santa Barbara, for providing the specifications of the heat-shock response model, and the members of the MOSAIC group (ETH Zurich) for fruitful discussions on the manuscript. We also thank the referee for the detailed comments, which greatly helped improving the manuscript, and Jo Helmuth (MOSAIC Group, ETH Zurich) for proofreading. RR thanks Omar Awile for his assistance in optimizing the implementation of PDM. RR was financed by a grant from the Swiss SystemsX.ch initiative, evaluated by the Swiss National Science Foundation. 

\appendix
\section{The original SSA algorithms}\label{app:ssa}
Gillespie's direct method (DM) consists of the following steps: 
\begin{enumerate}
\item Set \mbox{$t\,\leftarrow\,0$}; initialize $\vec{n}$, \mbox{$a_\mu \,\forall \mu$}, and $a$
\item Sample $\mu$: generate a uniform random number \mbox{$r_1 \in [0,1)$} and determine $\mu$ as the smallest integer satisfying \mbox{$r_1<\sum_{\mu '=1}^{\mu}a_{\mu '}/a$} (see Eq.~\ref{eq:jointpdf})
\item Sample $\tau$: generate a uniform random number \mbox{$r_2 \in [0,1)$} and compute the real number $\tau$ as \mbox{$\tau=-a^{-1}\ln(r_2)$} (see Eq.~\ref{eq:jointpdf})
\item Update: \mbox{$\vec{n}\,\leftarrow\,\vec{n}\,+\,{\boldsymbol \nu}_{\mu}$}, where \mbox{$\boldsymbol \nu_\mu$} is the stoichiometry of reaction $\mu$; recompute all \mbox{$a_\mu$} and $a$
\item \mbox{$t\,\leftarrow\,t\,+\,\tau$}; go to step 2
\end{enumerate} 

\medskip
The first reaction method (FRM) uses a different sampling strategy for $\mu$ and $\tau$ as follows: $\tau\,=\,\r{min}[\{\tau_1,\tau_2,\ldots,\tau_M\}]$ and $\mu$ is the index of the smallest $\tau$. The probability density of the time to the \mbox{$i^\text{th}$} reaction, \mbox{$\tau_i$}, is given by \mbox{$p_{\tau_i}\,=\,a_i\,\text{e}^{-a_i\tau_i}$}.

\section{Algebraic equivalence of PDM's sampling scheme to that of Gillespie's direct method}\label{pdm:proof}
In the direct method (DM), the next reaction index is sampled as
\be
\label{eq1:proof}
\mu ={\operatorname{min}}\left[ \mu ' \,\, : \,\, r_{1}a < \sum_{m=1}^{\mu '} a_{m} \right] \, , 
\ee
where $r_1$ is a uniform random number $\in\,[0,1)$ and $a_{m}$ is the propensity of reaction $m$. Without loss of generality, we identify $\mu '$ by a unique pair of indices, $I '$ and $J '$, such that $\mu '\,=\,\mathrm{L}_{I',J'}$. Using this mapping to a group (row) index $I'$ and an element (column) index $J'$, Eq.~\ref{eq1:proof} becomes 
\be
\label{eq2:proof}
\left(\begin{array}{c} I \\ J \end{array}\right) ={\operatorname{min}}\left[ \left(\begin{array}{c} I' \\ J' \end{array}\right) \,\, : \,\, r_{1}a < \sum_{i=0}^{I'-1}\sum_{\forall j} a_{\r{L}_{i,j}}\,+\,\sum_{j=1}^{J'} a_{\r{L}_{I',j}} \right] \, ,
\ee
such that $\mu\,=\,\r{L}_{I,J}$. 
This can be written for the group (row) index $I$ alone 
\be
\label{eq4:proof}
I ={\operatorname{min}}\left[ I' \,\, : \,\, r_{1}a < \sum_{i=0}^{I'}\sum_{\forall j} a_{\r{L}_{i,j}}\right]
\ee 
and the element (column) index $J$ alone
\be
\label{eq5:proof}
J ={\operatorname{min}}\left[ J' \,\, : \,\, r_{1}a < \sum_{i=0}^{I-1}\sum_{\forall j} a_{\r{L}_{i,j}} \,+\, \sum_{j=1}^{J'} a_{\r{L}_{I,j}}\right].
\ee 
Using the definitions for $\Sigma _i$ and ${\boldsymbol \Pi} _i$, Eqs.~\ref{eq4:proof} and~\ref{eq5:proof} are equivalent to Eqs.~\ref{row_mu} and~\ref{col_mu}, respectively.  

\section{Computational cost and memory requirement of PDM and SPDM}\label{app:complexity}

\subsection{Computational cost}\label{app:pdmtime}
The computational cost of PDM is governed by the following steps: (a) sampling the index of the next reaction and (b) updating the population $\vec{n}$ and the partial-propensity structure $\boldsymbol \Pi$. The computational cost of SPDM is the same as that of PDM.

\paragraph{Computational cost of sampling the index of the next reaction.} For any chemical reaction network with $N$ species, the number of arrays in the partial-propensity structure ${\boldsymbol \Pi}$ is at most $N+1$, which is also the maximum length of $\bsym \Sigma$ and $\bsym \Lambda$. The number of entries in each array \mbox{$\boldsymbol \Pi _{i}$} is at most \mbox{$2N$}, since any species can react with at most $N$ species in bimolecular reactions and undergo at most $N$ unimolecular reactions. Sampling the index of the next reaction involves two steps: (a) a linear search for the group index $I$ in $\bsym\Sigma$ and (b) a linear search for the element index $J$ in \mbox{$\bsym\Pi _I$}. Since $\bsym\Sigma$ is at most of length \mbox{$N+1$}, the first step is \mbox{$O(N)$}. The second step is also \mbox{$O(N)$}, since no \mbox{$\bsym\Pi _i$} can be longer than \mbox{$2N$}. The overall computational cost of sampling the next reaction is thus  \mbox{$O(N)$} for networks of any degree of coupling.

\paragraph{Computational cost of the update.} Let the maximum number of chemical species involved in any reaction (as reactants or products) be given by the constant \mbox{$s$} (constant with system size). The computational cost of updating $\vec{n}$ is thus \mbox{$s\in O(1)$}. In PDM, only the partial propensities of bimolecular reactions need to be updated. The total number of entries in the third update structure \mbox{$\mathrm{\mathbf{U}^{(3)}}$} is, thus, equal to the number of bimolecular reactions. In addition, the total number of entries in $\boldsymbol \Pi$ that depend on any \mbox{$n_i$} is always less than or equal to $N$, as any species S\mbox{$_i$} can only react with itself and the remaining \mbox{$N-1$} species in bimolecular reactions. Therefore, the upper bound for the total number of partial propensities in $\boldsymbol \Pi$ to be updated after executing any reaction is \mbox{$sN\in O(N)$}. \\    

In summary, the computational cost of PDM is \mbox{$O(N)$}, irrespective of the degree of coupling in the reaction network (see Table \ref{table_Cs} for benchmark results).   

\subsection{Memory requirement}\label{app:pdmspace}
The memory requirement of PDM is given by the total size of the data structures $\vec{n}$, $\boldsymbol \Pi$, \textbf{L}, $\boldsymbol \Lambda$, $\boldsymbol \Sigma$, \mbox{$\mathrm{\mathbf{U}^{(1)}}$}, \mbox{$\mathrm{\mathbf{U}^{(2)}}$}, and \mbox{$\mathrm{\mathbf{U}^{(3)}}$}. \\

The partial-propensity structure $\boldsymbol \Pi$ and the look-up table \textbf{L} have the same size. Since every reaction is accounted for exactly once, each structure requires $O(M)$ memory. $\boldsymbol \Lambda$, $\vec{n}$, and $\boldsymbol \Sigma$ are all at most of length \mbox{$N+1$} and thus require  \mbox{$O(N)$} memory. The sizes of \mbox{$\mathrm{\mathbf{U}^{(1)}}$} and \mbox{$\mathrm{\mathbf{U}^{(2)}}$} are \mbox{$O(M)$}, and the size of \mbox{$\mathrm{\mathbf{U}^{(3)}}$} is proportional the number of bimolecular reactions and, hence, \mbox{$O(M)$} if all reactions are bimolecular. \\

In summary, the memory requirement of PDM is \mbox{$O(M)$}. SPDM requires an additional \mbox{$N+M$} memory to store the reordered index lists (see Table \ref{table_Ms}).   

\section{Benchmark problem definitions}\label{app:testproblems}
\subsection{Colloidal aggregation model}\label{app:aggregation}
The reaction network of the colloidal aggregation model is defined by:
\bea
\notag
\r{S}_n + \r{S}_m \xrightarrow{c_{n,m}} & \r{S}_{n+m} 
\qquad & 
n=1,\ldots ,\left\lfloor \frac{N}{2}\right\rfloor \, ; \quad m=n,\ldots ,N-n  
\\
\label{Aggregation_reaction_equations2}
\r{S}_p \xrightarrow{\bar{c}_{p,q}} & \r{S}_q + \r{S}_{p-q} 
\qquad & 
p=1,\ldots ,N \,;\quad q=1,\ldots ,\left\lfloor \frac{p}{2}\right\rfloor \, . 
\eea
%

For an even number of species $N$, the partial-propensity structure for this network is: 
\bea
{\boldsymbol \Pi} =\left\{\begin{array}{l}
{\boldsymbol \Pi_0}\,=\,(\emptyset)\\
{\boldsymbol \Pi_1} \,=\,\left(c_{1,1}\frac{n_1-1}{2}\quad c_{1,2}n_2\quad c_{1,3}n_3\quad\dots\quad c_{1,\frac{N}{2}}n_{\frac{N}{2}}\quad\dots\quad c_{1,N-1}n_{N-1}\right)\\
{\boldsymbol \Pi_2}\,=\,\left(\bar{c}_{2,1}\quad c_{2,2}\frac{n_2-1}{2}\quad c_{2,3}n_3\quad\dots\quad c_{2,\frac{N}{2}}n_{\frac{N}{2}}\quad\dots\quad c_{2,N-2}n_{N-2}\right)\\
\vdots\\
{\boldsymbol \Pi_{\frac{N}{2}}}\,=\,\left(\bar{c}_{\frac{N}{2},1}\quad\bar{c}_{\frac{N}{2},2}\quad\dots\quad\bar{c}_{\frac{N}{2},\frac{N}{4}}\quad c_{\frac{N}{2},\frac{N}{2}}n_{\frac{N}{2}}\right)\\
\vdots\\
{\boldsymbol \Pi_N}\,=\,\left(\bar{c}_{N,1}\quad\bar{c}_{N,2}\quad\bar{c}_{N,3}\quad\dots\quad\ldots\quad\bar{c}_{\frac{N}{2},\frac{N}{2}}\right) \, . 
\end{array}\right.
\label{eq:Pi}
\eea
For odd $N$, the structure looks similar.\\ 

\subsection{Network of bimolecular reactions}\label{app:tightly}
The network of bimolecular reactions is given by:
\bea
\label{Tightly_coupled_reaction_equations}
\r{S}_n + \r{S}_m \xrightarrow{c_{n,m}} \r{S}_{p} + \r{S}_{q} \qquad \left\{\begin{array}{l}
n=1,\ldots ,N-1 \,;\quad m=n+1,\ldots ,N; \\
p=\r{min}\left[\{1,\ldots,N\}\backslash\{n,m\}\right]\,;\quad q=\r{min}\left[\{1,\ldots,N\}\backslash\{n,m,p\}\right] \, . 
\end{array}\right.
\eea

The partial-propensity structure for this reaction network is: 
\bea
{\boldsymbol \Pi} = \left\{\begin{array}{l}
{\boldsymbol \Pi_0}\,=\,(\emptyset)\\
{\boldsymbol \Pi_1} \,=\,\left(c_{1,2}n_2\quad c_{1,3}n_3\quad c_{1,4}n_4\quad\dots\quad c_{1,N}n_N\right)\\
{\boldsymbol \Pi_2}\,=\,\left(c_{2,3}n_3\quad c_{2,4}n_4\quad c_{2,5}n_5\quad\dots\quad c_{2,N}n_N\right)\\
\vdots\\
{\boldsymbol \Pi_{N-1}}\,=\,\left(c_{N-1,N}n_N\right)\\
{\boldsymbol \Pi_{N}}\,=\,\left(\emptyset\right) \, . 
\end{array}\right.
\label{eq_tightly_coupled_rn:Pi}
\eea 

\subsection{Linear chain model}\label{app:linearchain}
The reactions of the linear chain model are given by:
\bea
\r{S}_i \xrightarrow{c_{i}} & \r{S}_{i+1} 
\qquad & 
i=1,\ldots ,N-1 \, ,
\label{linear_chain_reaction_equations}
\eea

and the partial-propensity structure is: 
\bea
{\boldsymbol \Pi} = \left\{\begin{array}{l}
{\boldsymbol \Pi_0}\,=\,(\emptyset)\\
{\boldsymbol \Pi_1} \,=\,(c_1)\\
{\boldsymbol \Pi_2}\,=\,(c_2)\\
\vdots\\
{\boldsymbol \Pi_{N-1}}\,=\,(c_{N-1})\\
{\boldsymbol \Pi_{N}}\,=\,(\emptyset) \, . 
\end{array}\right.
\label{eq_linear_model:Pi}
\eea

\subsection{Heat-shock response model}
The heat-shock response model \cite{Kurata:2001} was obtained from Dr.~Hong Li and Prof.~Linda Petzold (UCSB) and is publicly available as part of the StochKit package \cite{Li:2008}.

\newpage

\section*{Figures}

\begin{figure}[htbp]
\begin{center}
\subfigure[(a) Colloidal aggregation model]{{\label{Aggregation_model1}}\epsfig{file=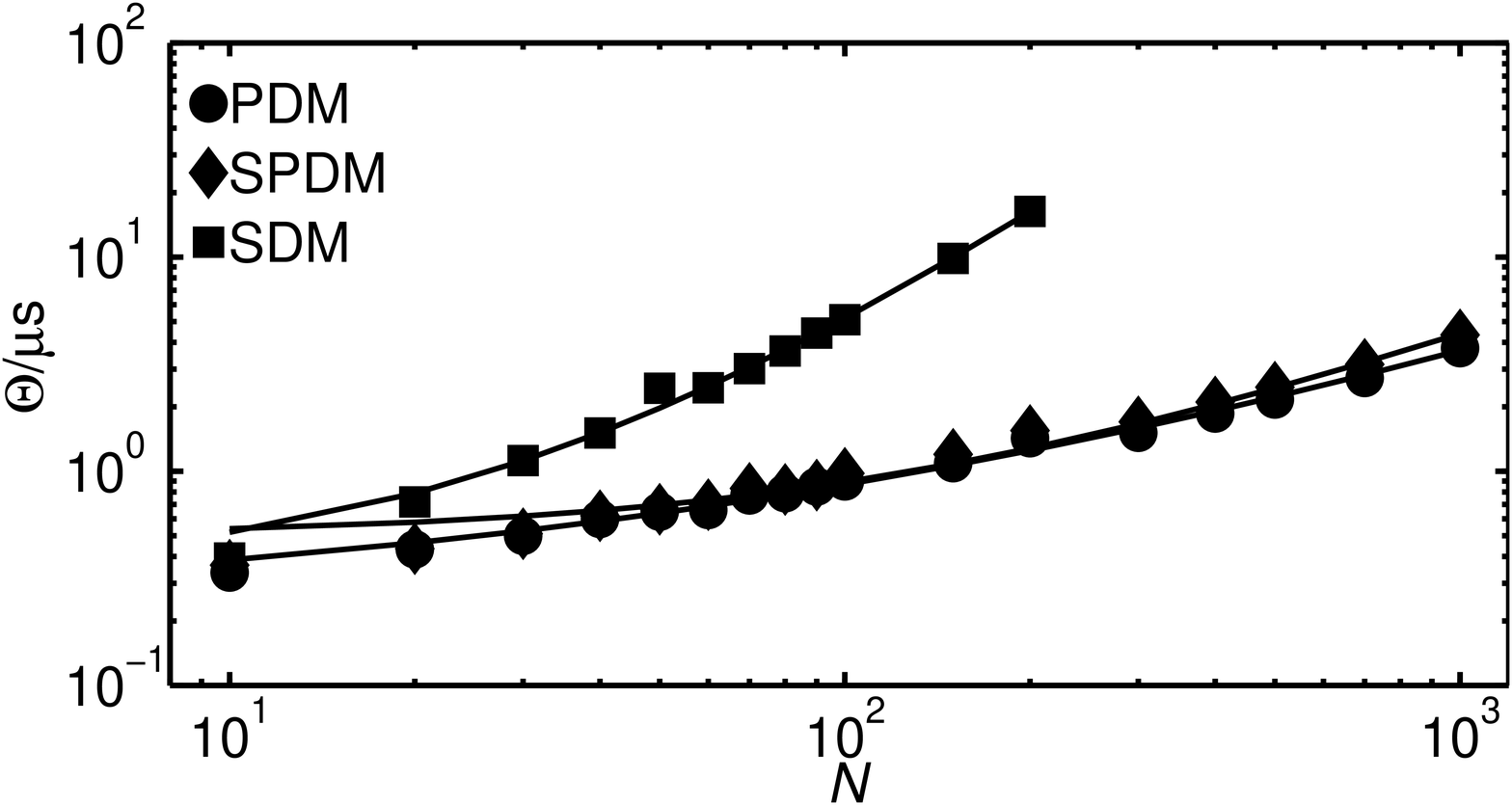, angle=0, height=4.4cm}}
\subfigure[(b) Network of bimolecular reactions]{{\label{Tightly_coupled_model}}\epsfig{file=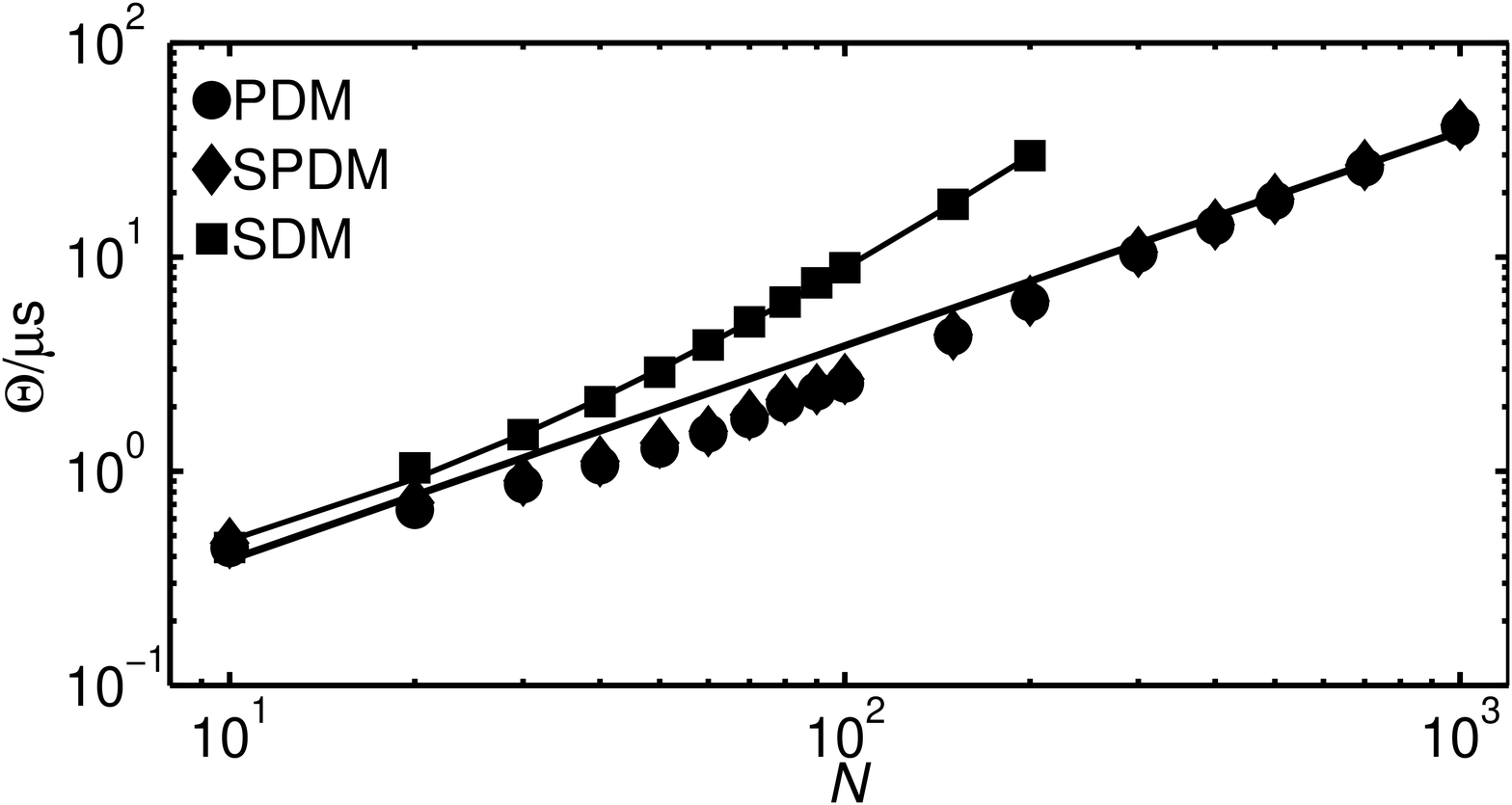, angle=0, height=4.4cm}}
\subfigure[(c) Linear chain model]{{\label{Linear_model1}}\epsfig{file=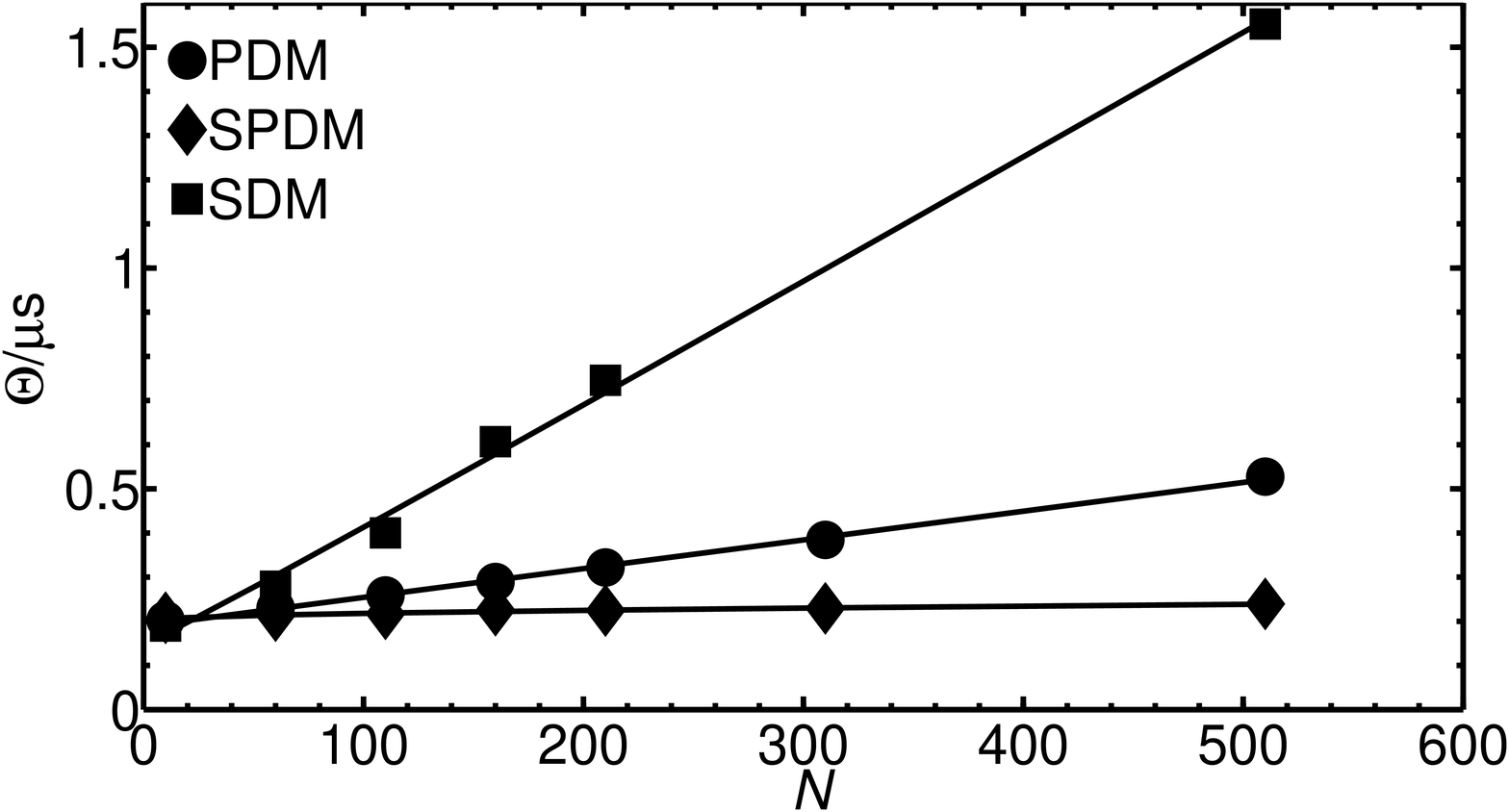, angle=0, width=8.3cm,height=4.4cm}}
\caption{Computational costs of PDM (circles), SPDM (diamonds), and SDM (squares). See main text for the simulation parameters and initial conditions used. The average CPU time per reaction (i.e.~per time step), $\Theta$, is shown as a function of system size quantified by the number of species $N$.
$\Theta$ is defined as the CPU time needed to simulate the system up to final time $T$, divided by the number of reactions executed during this time, and averaged over 100 independent runs (error bars are smaller than symbol size). The solid lines are the corresponding least-squares fits of the scaling $\Theta(N)$ of PDM, SPDM, and SDM with the model \mbox{$a\mathcal{C}_\mu+b\mathcal{C}_\mathbf{n}+c\mathcal{C}_\r{P}$} on a linear scale (see Table \ref{table_Cs}), where $a$, $b$, and $c$ are the fitted constants. (a) Logarithmic plot of the results for the colloidal aggregation model. The fits are: \mbox{$\Theta^\text{PDM}/\mathrm{\mu s}$} = \mbox{$0.0022N+0.050N^{0.5}+0.22$}, \mbox{$\Theta^\text{SPDM}/\mathrm{\mu s}$} = \mbox{$0.0027N+0.053N^{0.5}+0.20$}, and \mbox{$\Theta^\text{SDM}/\mathrm{\mu s}$} = \mbox{0.00031$N^{2}+$0.018$N+$0.31}. (b) Logarithmic plot of the results for the network of bimolecular reactions. The fits are: \mbox{$\Theta^\text{PDM}/\mathrm{\mu s}$} = \mbox{0.038$N$}, \mbox{$\Theta^\text{SPDM}/\mathrm{\mu s}$} = \mbox{0.039$N$}, and \mbox{$\Theta^\text{SDM}/\mathrm{\mu s}$} = \mbox{0.00061$N^{2} +$0.027$N+$0.15}. (c) Linear plot of the results for the linear chain model. The fits are: \mbox{$\Theta^\text{PDM}/\mathrm{\mu s}$} = \mbox{0.00065$N+$0.19}, \mbox{$\Theta^\text{SPDM}/\mathrm{\mu s}$} = \mbox{0.0015$N^{0.5}+$0.20}, and \mbox{$\Theta^\text{SDM}/\mathrm{\mu s}$} = \mbox{0.0029$N-$0.0025$N^{0.5}+$0.15}. In all cases, the computational cost $\Theta (N)$ of PDM and SPDM is $O(N)$.}
\label{Theta_vs_N_plot}\label{Aggregation_model1}\label{Tightly_coupled_model}\label{Linear_model1}
\end{center}
\end{figure}

\newpage

\section*{Tables}
\begin{table}[htbp]
\begin{itemize}
\item[1.] Initialization: set \mbox{$t\,\leftarrow\,0$}; initialize $\vec{n}$, $\boldsymbol \Pi$, $\boldsymbol \Lambda$, $\boldsymbol \Sigma$; $a\,\leftarrow\,\sum_{i=0}^{N}\Sigma_i$; \mbox{$\Delta a\,\leftarrow\,0$}; generate $\mathrm{\mathbf{L}}$, $\mathrm{\mathbf{U}^{(1)}}$, $\mathrm{\mathbf{U}^{(2)}}$, and $\mathrm{\mathbf{U}^{(3)}}$

\item[2.] Sample $\mu$: generate a uniform random number $r_1\in [0,1)$ and determine the group index $I$ and the element index $J$ according to Eqs.~(\ref{row_mu}),~(\ref{col_mu1}), and~(\ref{col_mu3}); \mbox{$\mu\,\leftarrow\,\r{L}_{I,J}$}

\item[3.] Sample $\tau$: generate a uniform random number $r_2\in [0,1)$ and compute the time to next reaction $\tau$ as \mbox{$\tau\,\leftarrow\,a^{-1}\ln(r_2^{-1})$} 

\item[4.]Update $\vec{n}$: for each index $k$ of $\mathrm{\mathbf{U}^{(1)}_\mu}$, \mbox{$l\,\leftarrow\,\mathrm{U}^{(1)}_{\mu,k}$} and \mbox{$n_{l}\,\leftarrow\,n_{l}\,+\,\mathrm{U}^{(2)}_{\mu,k}$} 

\item[5.] Update $\boldsymbol \Pi$, $\boldsymbol \Lambda$, $\boldsymbol \Sigma$ and compute $\Delta a$, the change in $a$: 
\item[] For each index $k$ of $\mathrm{\mathbf{U}^{(1)}_\mu}$, do:

\bi
\item[5.1.] \mbox{$l\,\leftarrow\,\mathrm{U}^{(1)}_{\mu,k}$}

\item[5.2.] For each index $m$ of $\mathrm{\mathbf{U}}^{(3)}_{l}$, do:

\bi
\item[5.2.1.] \mbox{$(i_m^l,\, j_m^l)\,\leftarrow \, \mathrm{U}^{(3)}_{l,m}$} \quad(Eq.~\ref{U3})
\item[5.2.2.] \mbox{$\Pi_{i_{m}^{l},j_{m}^{l}}\,\leftarrow\,\Pi_{i_{m}^{l},j_{m}^{l}}\,+\,c_\mu\mathrm{U}^{(2)}_{\mu,k}$}, \quad if $l$$\neq$$i_m^l$\\
\mbox{$\Pi_{i_{m}^{l},j_{m}^{l}}\,\leftarrow\,\Pi_{i_{m}^{l},j_{m}^{l}}\,+\,\frac{1}{2}c_\mu\mathrm{U}^{(2)}_{\mu,k}$}, \quad if $l$=$i_m^l$

\item[5.2.3.] \mbox{$\Lambda_{i_{m}^{l}}\,\leftarrow\,\Lambda_{i_{m}^{l}}\,+\,c_\mu\mathrm{U}^{(2)}_{\mu,k}$}, \quad if $l$$\neq$$i_m^l$\\ 
\mbox{$\Lambda_{i_{m}^{l}}\,\leftarrow\,\Lambda_{i_{m}^{l}}\,+\,\frac{1}{2}c_\mu\mathrm{U}^{(2)}_{\mu,k}$}, \quad if $l$=$i_m^l$

\item[5.2.4.] \mbox{$\Sigma_\mathrm{temp}\,\leftarrow\,\Sigma_{i_{m}^{l}}$}

\item[5.2.5.] \mbox{$\Sigma_{i_{m}^{l}}\,\leftarrow\,n_{i_{m}^{l}} \Lambda_{i_{m}^{l}}$}

\item[5.2.6.] \mbox{$\Delta a\,\leftarrow\,\Delta a\,+\,\Sigma_{i_{m}^{l}}\,-\,\Sigma_\mathrm{temp}$}
\ei

\item[5.3.] \mbox{$\Delta a\,\leftarrow\,\Delta a\,+\,n_l\Lambda_l\,-\,\Sigma_l$}; \,\,$\Sigma_l\,\leftarrow\,n_l\Lambda_l$

\ei

\item[6.] Update $a$ and increment time: \mbox{$a\,\leftarrow\,a\,+\,\Delta a$}; \,\,\mbox{$\Delta a\,\leftarrow\,0$}; \,\,\mbox{$t\,\leftarrow\,t\,+\,\tau$}

\item[7.] Go to step 2

\end{itemize}
\caption{Detailed algorithm for the partial-propensity direct method PDM.}
\label{our_algorithm}
\end{table}

\begin{table}[htbp]
\renewcommand{\arraystretch}{1.4}
\begin{tabular}{|c|c|c|c|c|c|c|}
\hline
Model & Number of & Number of &  \multicolumn{3}{|c|}{Degree of coupling ($k$)}\\
\cline{4-6}
	 & species ($N$) & reactions ($M$) & Minimum & Average & Maximum\\
\hline
CA & $N$ & $\left\lfloor\frac{N^2}{2}\right\rfloor$ & $1.3N-0.33$ & $2.3N-4.7$ & $3.0N-7.0$\\
NB & $N$ & $\frac{N(N-1)}{2}$ & $4.0N-10$ & $4.0N-10$ & $4.0N-10$\\
LC & $N$ & $N-1$ & 1$^{(*)}$ & $2-\frac{1}{N-1}\approx2$ & 2\\
HSR & 28 & 61 & 1 & 5.9 & 11\\
\hline
\end{tabular}
\caption{Properties of the benchmark cases. The number of species, number of reactions, and minimum, average, maximum out-degree of the dependency graph (degree of coupling) are given for the benchmark cases defined in Appendix \ref{app:testproblems}: the colloidal aggregation model (CA), the network of bimolecular reactions (NB), the linear chain model (LC), and the heat-shock response model (HSR). ($^{*}$) In the linear chain model the degree of coupling is 1 only for the last reaction, since its product is not a reactant anywhere else.}
\label{table_test_cases}
\end{table}

\begin{table}[htbp]
\renewcommand{\arraystretch}{1.2}
\begin{tabular}{|c|c|c|c|c|c|c|}
\cline{1-7}
\multirow{2}{*}{} & \multicolumn{3}{|c|}{PDM}& \multicolumn{3}{|c|}{SPDM}\\
\cline{2-7}
& $\mathcal{C}_\mu$ & $\mathcal{C}_\mathbf{n}$ & $\mathcal{C}_\r{P}$ & $\mathcal{C}_\mu$ & $\mathcal{C}_\mathbf{n}$ & $\mathcal{C}_\r{P}$\\
\cline{1-7}
CA & $0.49N+2.0$ & 3 & $5.2N^{0.5}-8.1$ & $0.45N+0.38$ & 3 & $5.2N^{0.5}-8.1$\\
NB & $0.97N-1.3$ & 4 & $1.6N-3.2$ & $0.94N-4.7$ & 4 & $1.6N-3.2$\\ 
LC & $0.50N+1.0$ & 2 & 0 & $1.0N^{0.5}+0.79$ & 2 & 0\\
HSR & 13 & 3 & 2.2 & 3.7 & 3 & 2.2\\
\cline{1-7}
\end{tabular}
\renewcommand{\arraystretch}{1.2}
\begin{tabular}{|c|c|c|c|}
\cline{1-4}
\multirow{2}{*}{} & \multicolumn{3}{|c|}{SDM}\\
\cline{2-4}
& $\mathcal{C}_\mu$ & $\mathcal{C}_\mathbf{n}$ & $\mathcal{C}_\r{P}$\\
\cline{1-4}
CA & $0.14N^2+1.2N-9.9$ & $N$ & $2.8N-10$\\
NB & $0.33N^2-0.044N+0.51$ & $N$ & $4.0N-10$\\ 
LC & $1.0N^{0.5}-0.21$ & $N$ & 2\\
HSR & 2.9 & 28 & 8.2\\
\cline{1-4}
\end{tabular}
\caption{Number of compute operations needed by the different algorithms (PDM, SPDM, SDM) for the different test cases (CA: colloidal aggregation model; NB: network of bimolecular reactions; LC: linear chain model; HSR: heat-shock response model). \mbox{$\mathcal{C}_\mu$} is the average number of operations needed to sample the next reaction $\mu$. \mbox{$\mathcal{C}_\mathbf{n}$} is the average number of entries in the population $\vec{n}$ that need to be updated after any reaction. \mbox{$\mathcal{C}_\r{P}$} is the average number of partial propensities (or propensities for SDM) that need to be updated after any reaction. The operation counts are averaged over all reactions executed during 100 independent runs of each benchmark over the range of $N$ shown in Fig.~\ref{Theta_vs_N_plot}. The average numbers are then fitted with the models given here (with correlation coefficient of at least 0.98 in all cases). See Fig.~\ref{Upsilon_vs_N_plot} for the distribution of the number of updates.}
\label{table_Cs}
\end{table}

\begin{table}[htbp]
\renewcommand{\arraystretch}{1.4}
\begin{tabular}{|c|c|c|c|c|c|}
\cline{1-6}
\multirow{2}{*}{} & \multicolumn{5}{|c|}{PDM/SPDM}\\
\cline{2-6}
& $\mathbf{n}$, ${\boldsymbol \Lambda}$, ${\boldsymbol \Sigma}$ & ${\boldsymbol \Pi}$, $\mathrm{\mathbf{L}}$, $\mathbf{c}$ & $\mathrm{\mathbf{U}}^{(1)}$, $\mathrm{\mathbf{U}}^{(2)}$ &  $\mathrm{\mathbf{U}}^{(3)}$ & Total\\
\hline
CA & $N$ & $\left\lfloor\frac{N^2}{2}\right\rfloor$ & 3$\left\lfloor\frac{N^2}{2}\right\rfloor$ & 2$\left\lfloor\frac{N^2}{4}\right\rfloor$ & $O(N^2)\,=\,O(M)$\\
NB & $N$ & $\frac{N(N-1)}{2}$ & $4\frac{N(N-1)}{2}$ & $2\frac{N(N-1)}{2}$ & $O(N^2)\,=\,O(M)$\\ 
LC & $N$ & $N-1$ & $2(N-1)$ & 0 & $O(N)\,=\,O(M)$\\  
HSR & 28 & 61 & 133 & 24 & 557\\
\hline
\end{tabular}\\
\renewcommand{\arraystretch}{1.4}
\begin{tabular}{|c|c|c|c|c|c|}
\hline
\multirow{2}{*}{} & \multicolumn{5}{|c|}{SDM}\\
\cline{2-6}
& $\mathbf{n}$ & $\mathbf{c}$, $\mathbf{a}$ & dependency graph & ${\boldsymbol \nu}$ & Total\\
\hline
CA & $N$ & $\left\lfloor\frac{N^2}{2}\right\rfloor$ & $1.2N^3-2.5N^2+2.3N$ & $N\left\lfloor\frac{N^2}{2}\right\rfloor$ & $O(N^3)\,=\,O(NM)$\\
NB & $N$ & $\frac{N(N-1)}{2}$ & $2N^3-7N^2+5N$ & $\frac{N^2(N-1)}{2}$ & $O(N^3)\,=\,O(NM)$\\ 
LC & $N$ & $N-1$ & $2(N-1)$ & $N(N-1)$ & $O(N^2)\,=\,O(NM)$\\  
HSR & 28 & 61 & 360 & 1708 & 2218\\
\hline
\end{tabular}
\caption{Total amount of computer memory needed by the different algorithms (PDM, SPDM, SDM) for the different test cases (CA: colloidal aggregation model; NB: network of bimolecular reactions; LC: linear chain model; HSR: heat-shock response model). The sizes of all major data structures (\mbox{$\mathbf{c}$} and \mbox{$\mathbf{a}$} are the arrays of specific probability rates and reaction propensities, respectively;  $\boldsymbol \nu$ is the stoichiometry matrix; see Sec.~\ref{sec:methods_description}\ref{sec:PDM_description} for other definitions) as well as the total memory requirements are given as determined analytically for all benchmark simulations. SPDM and SDM need additional memory of size \mbox{$M+N$} and \mbox{$M$}, respectively, for the reordered index lists. This, however, does not change the overall scaling of the total memory requirements.}
\label{table_Ms}
\end{table}

\end{document}